\documentclass[preprint,notoc]{JHEP3}
\usepackage{epsfig}

\def\eslt{E_T^{\rm miss}}

\def\to{\rightarrow}

\def\bi{\begin{itemize}}
 \def\ei{\end{itemize}}
\def\te{\tilde e}

\def\c1p{C1^\prime}
\def\msq3{\overline{m}_{\tilde{q}}(3)}

\def\tu{\tilde u}

\def\tb{\tilde b}

\def\tst{\tilde t}
\def\ttau{\tilde \tau}

\def\tg{\tilde g}
\def\tnu{\tilde\nu}
\def\tell{\tilde\ell}
\def\tq{\tilde q}

\def\tw{\widetilde W}
\def\tz{\widetilde Z}
\def\alt{\lesssim}
\def\agt{\gtrsim}
\def\be{\begin{equation}}  
\def\ee{\end{equation}}  
\def\bea{\begin{eqnarray}}  
\def\eea{\end{eqnarray}}  
\def\CM{{\cal{M}}}
\def\sps1ap{SPS1a$^\prime$}

\newcommand\epjh[3]{{\it Eur.\ Phys.\ J. \ }{\bf H #1} (#2) #3}
\newcommand\jpcs[3]{{\it J. \ Phys. \ Conf. \ Ser.\ }{\bf #1} (#2) #3}

\title{Post-LHC7 fine-tuning 
in the mSUGRA/CMSSM model with a 125~GeV Higgs boson
}
\author{Howard Baer$^{a}$, Vernon Barger$^b$, Peisi Huang$^b$, 
Dan Mickelson$^{a}$, Azar Mustafayev$^c$ and Xerxes Tata$^c$\\
$^a$Dept. of Physics and Astronomy, University of Oklahoma, Norman, OK 73019, USA\\
$^b$Dept. of Physics, University of Wisconsin, Madison, WI 53706, USA\\
$^c$Dept. of Physics and Astronomy, University of Hawaii, Honolulu, HI 96822, USA\\
E-mail: \email{baer@nhn.ou.edu}, \email{barger@pheno.wisc.edu},
\email{phuang7@wisc.edu}, \email{dsmickelson@ou.edu}, 
\email{azar@phys.hawaii.edu}, \email{tata@phys.hawaii.edu}
}

\preprint{\vbox{UH-511-1202-12}}

\abstract{The recent discovery of a 125~GeV Higgs-like resonance at
LHC, coupled with the lack of evidence for weak scale supersymmetry
(SUSY), have severely constrained SUSY models such as mSUGRA/CMSSM. As
LHC probes deeper into SUSY model parameter space, the little hierarchy
problem -- how to reconcile the $Z$ and Higgs boson mass scale with the
scale of SUSY breaking -- will become increasingly exacerbated unless a
sparticle signal is found.  We evaluate two different measures of
fine-tuning in the mSUGRA/CMSSM model. The more stringent of these,
$\Delta_{\rm HS}$, includes effects that arise from the high scale
origin of the mSUGRA parameters while the second measure, $\Delta_{\rm
EW}$, is determined only by weak scale parameters: hence, it is universal
to any model with the same particle spectrum and couplings. Our results
incorporate the latest constraints from LHC7 sparticle searches, LHCb limits
from $B_s\to\mu^+\mu^-$ and also
require a light Higgs scalar with $m_h\sim 123-127$~GeV.  We present
fine-tuning contours in the $m_0\ vs.\ m_{1/2}$ plane for several sets
of $A_0$ and $\tan\beta$ values. We also present results for
$\Delta_{\rm HS}$ and $\Delta_{\rm EW}$ from a scan over the entire
viable model parameter space.  We find a $\Delta_{\rm HS}\agt 10^3$, or
at best $0.1\%$ fine-tuning.  For the less stringent electroweak fine
tuning, we find $\Delta_{\rm EW}\agt 10^2$, or at best 1\% fine-tuning.
Two benchmark points are presented that have the lowest values of
$\Delta_{\rm HS}$ and $\Delta_{\rm EW}$. 
Our results provide a
quantitative measure for ascertaining whether or not the remaining 
mSUGRA/CMSSM model parameter space is excessively fine-tuned, and so
could provide impetus for considering alternative SUSY models.
}
\begin{document}

\section{Introduction}
\label{sec:intro}

The recent spectacular runs of LHC at $\sqrt{s}=7$ and 8~TeV have led to
identification of a Higgs-like boson\footnote{This particle has spin 0
or $\geq 2$ and couples directly to the $ZZ$, and with weaker evidence
also to the $WW$,
systems. The latter property implies a connection with electroweak
symmetry breaking, characteristic of the Higgs boson.} with mass
$m_h\sim 125$~GeV~\cite{atlas_h,cms_h}.  This is in accord with
predictions from the minimal supersymmetric standard model (MSSM) which
requires that the lighter higgs scalar mass $m_h\alt
130-135$~GeV~\cite{mh}. Since values of $m_h > M_Z$ are only possible
due to radiative corrections, the upper end of the range depends on the
masses of third generation sparticles that one is willing to allow.  To
achieve $m_h\sim 125$~GeV, either large mixing or several TeV masses are
required in the top squark sector.  In models such as the much-studied
minimal supergravity (mSUGRA or CMSSM) model~\cite{msugra,kkrw}, values
of trilinear soft breaking parameter $|A_0|\sim (1.5-2)m_0$ are favored,
along with top squark masses $m_{\tst_{1,2}}\agt 1-2$~TeV: for positive
$A_0$ values $m_0$ is typically larger than 5~TeV~\cite{bbm,others}.

While the measured value of $m_h$ is within the expected range of even
the simplest SUSY models,
there is at present no sign of SUSY particles at LHC. From LHC data
analyses within the mSUGRA model, mass limits of $m_{\tg}\agt 1.4$~TeV
when $m_{\tq}\sim m_{\tg}$ and $m_{\tg}\agt 0.9$~TeV when $m_{\tq}\gg
m_{\tg}$ have been reported\cite{atlas_susy,cms_susy}.  Several groups~\cite{cmssm}
have updated their fits of the mSUGRA/CMSSM model to various data sets,
now including information from LHC7 and LHC8 Higgs-like boson discovery  
and LHC7 sparticle mass limits.  Typically, the best fit
regions have moved out to  large values of $m_0$ and $m_{1/2}$ to
accomodate the LHC sparticle mass limits and Higgs discovery. 
Such large $m_0$ and $m_{1/2}$ values lead to sparticle masses in the
multi-TeV mass range, thus exacerbating what has become known as {\it
the little hierarchy problem}: how do such large SUSY particle masses
and soft breaking parameters conspire to yield the weak scale typified
by the $Z$-boson mass $M_Z\simeq 91.2$~GeV. The conflict between the
strong new LHC sparticle mass limits and the comparatively low values of
$M_Z$ and $m_h$ has intensified interest in the 
fine-tuning in supersymmetric models\cite{BG,kn,ftpapers,bbhmt,many}.

To set the stage for this analysis, we begin by reviewing radiative
corrections (assumed perturbative) to scalar field masses. In a {\it
generic} quantum field theory, taken to be the low energy effective
theory whose domain of validity extends up to the energy scale
$\Lambda$, the physical mass squared of scalar fields takes the
schematic form (at leading order),
\begin{equation}
m_{\phi}^2 = m_{\phi 0}^2 + C_1 \frac{g^2}{16\pi^2}\Lambda^2 + 
C_2 \frac{g^2}{16\pi^2}m_{\rm low}^2 \log\left(\frac{\Lambda^2}{m_{\rm low}^2}\right) 
+C_3 \frac{g^2}{16 \pi^2}m_{\rm low}^2.
\label{eq:generic}
\end{equation}
In Eq.~(\ref{eq:generic}), $g$ denotes the typical coupling of the
scalar $\phi$, $m_{\phi 0}$ is the corresponding mass parameter in the
Lagrangian, $16\pi^2$ is a loop factor, and $C_i$ are constants that aside
from spin, colour and other multiplicity factors are numbers 
${\cal O}(1)$. 
The scales $m_{\rm low}$ and $\Lambda$ respectively denote the 
highest mass scale in the effective theory and the scale at which this
effective theory description becomes invalid because heavy degrees of
freedom not included in the low energy Lagrangian become
important.  For instance, if we are considering corrections to the 
Higgs sector of the MSSM is embedded into a Grand Unified
Theory (GUT) framework, $\Lambda \sim M_{\rm GUT}$ and $m_{\rm low} \sim
M_{\rm SUSY}$ (or more precisely, $m_{\rm low}$ is around the mass of
the heaviest sparticles that have large couplings to the scalar
$\phi$). Finally, the  last term  in (\ref{eq:generic}) comes from loops of
particles of the low energy theory, and their scale is set by $m_{\rm
  low}$. These terms may contain logarithms, but {\it no large
  logarithms} since effects of very high momentum loops are included in
the $C_1$ and $C_2$ terms. These finite corrections provide
contributions to that which we have referred to as electroweak fine-tuning in
a previous study\cite{bbhmt}.

If the effective theory description is assumed to be valid to the GUT
scale, the $C_1$ term is enormous.  Even so it is always possible
to adjust the Lagrangian parameter $m_{\phi 0}^2$ to get the desired
value of $m_{\phi^2}\alt m_{\rm low}^2$. This is the {\it big fine
tuning problem} of generic quantum field theory with elementary
scalars. This problem is absent in softly broken supersymmetric theories
because $C_1 =0$. We see from Eq.~(\ref{eq:generic}) that if the
physical value of $m_{\phi}$ is significantly smaller than $m_{\rm low}$
(which in the case of the MSSM $\sim m_{\tst_i}$), we will still need to
have significant cancellations among the various terms to get the
desired value of $m_{\phi}$. This is the {\em little hierarchy problem.}
We also see that in models such as mSUGRA
that are assumed to be valid up to very high energy scales $\Lambda \sim
M_{\rm GUT}-M_P$, the magnitude of the $C_2$ term typically far exceeds that 
of the $C_3$ term because the logarithm is large, and hence is
potentially the largest source of fine-tuning in such SUSY scenarios. 

Because the $C_2$ and $C_3$ terms in Eq.~(\ref{eq:generic}) have somewhat
different origins -- the $C_2$ term represents corrections from physics
at scales between $m_{\rm low}$ and $\Lambda$, while the $C_3$ term
captures the corrections from physics at or below the scale $m_{\rm
low}$ -- we will keep individual track of these terms. 
In the following we will refer to fine-tuning from $C_2$ type terms as
high scale fine-tuning (HSFT) (since this exists only in models that are
valid to energy scales much larger than $m_{\rm low}$) and to the fine
tuning from $C_3$-type terms as electroweak fine-tuning (EWFT) for
reasons that are evident. 
We emphasize that the sharp
distinction between these terms exists only in models such as mSUGRA
that are assumed to be a valid description to very high scales, and is
absent in low scale models such as the phenomenological MSSM~\cite{pmssm}. 

In this paper, we quantify the severity of fine-tuning in the mSUGRA
model, keeping separate
the contributions from the two different terms. We are motivated to do
so for two different reasons.
\begin{itemize}
\item First, as emphasized, $C_2$ type terms appear only if the theory
  is applicable out to scale $\Lambda \gg m_{\rm low}$, while the $C_3$
  type terms are always present. In this sense, the fine-tuning from the
  $C_3$ type terms is ubiquitous to all models, whereas the fine-tuning
  associated with the (potentially larger) $C_2$ type terms may be
  absent, depending on the model. 

\item Second, as we will explain below, there are two very
  different attitudes that one can adopt for the fine-tuning from $C_2$ type
  terms. Keeping the contributions from $C_2$ and $C_3$ separate will
  allow the reader the choice as to how to interpret our results and
  facilitate connection with previous studies.

\end{itemize}
The remainder of this paper is organized as follows.  In
Sec.~\ref{sec:FT} we introduce our measures of fine-tuning. As usual, we
adopt the degree to which various contributions from the minimization of
the one-loop effective potential in the MSSM Higgs boson sector must
cancel to reproduce the observed value of $M_Z^2$ as our measure of fine
tuning. We use these considerations to introduce two different
measures. The first of these is the less stringent one and relies only
on the weak scale Lagrangian that arises from mSUGRA with total
disregard for its high scale origin, and is referred to as electroweak
fine-tuning (EWFT). The other measure that we introduce incorporates the
high scale origin of mSUGRA parameters and is therefore referred to as
high scale fine-tuning (HSFT).  In Sec.~\ref{sec:plane}, we present
contours for both HSFT and EWFT in several mSUGRA $m_0\ vs.\ m_{1/2}$
planes along with excluded regions from LHC7 sparticle searches and
LHCb limits from $B_s\to\mu^+\mu^-$ searches.\footnote{We note that Z-pole observables such as
$A^b_{FB}$~\cite{pdg} and, according to recent calculation\cite{freitas} also 
$R_b\equiv\frac{\Gamma (Z\to b\bar{b})}{\Gamma (Z\to all)}$, 
appear to exhibit deviations at the $(2-2.5)\ \sigma$ level from Standard Model expectations. 
While these possible discrepancies merit a watchful eye, 
an attempt to account for them in a SUSY framework is beyond the scope of this paper.}
We find that while LHC7 sparticle mass limits typically require EWFT at
$\sim 1\%$ level, the requirement that $m_h\sim 125$~GeV 
leads to 
much more severe EWFT in the 0.1\% range in the bulk of parameter space. As
anticipated, HSFT is even more severe.  We also find that the hyperbolic
branch/focus point region (HB/FP)\cite{hb_fp} -- while enjoying lower
EWFT than the bulk of mSUGRA parameter space -- still requires
fine-tuning at about the percent level.  The fine-tuning situation is
exacerbated by the requirement of large $|A_0/m_0|$ for which the HB/FP
region is absent, resulting in large EWFT (and even larger HSFT).  In
Sec.~\ref{sec:scan}, we present results from a complete scan over
mSUGRA/CMSSM parameter space. In this case, respecting both the LHC7
sparticle mass bounds, LHCb results on $B_s\to\mu^+\mu^-$ and 
$m_h=123-127$~GeV (in accord with the
estimated theory error on our calculation of $m_h$), we find parameter
space points with maximally 0.1\% HSFT and 1\% EWFT.  We leave it to the
reader to assess how much fine-tuning is too much, and also to judge the
role of HSFT in models such as mSUGRA/CMSSM that originate in high scale
physics. We present and qualitatively discuss the phenomenology of two
model points with the lowest HSFT and the lowest EWFT in
Sec.~\ref{sec:bm}. We end with some concluding remarks and our
perspective in Sec.~\ref{sec:conclude}.

\section{Fine-tuning}
\label{sec:FT}

We begin by first writing the Higgs potential whose minimization
determines the electroweak gauge boson masses as, 
\bea
V_{Higgs}&=&(m_{H_u}^2+\mu^2)|h_u^0|^2 +(m_{H_d}^2+\mu^2)|h_d^0|^2
\nonumber \\ 
&&-B\mu (h_u^0h_d^0+h.c.)+{1\over 8}(g^2+g^{\prime 2})
(|h_u^0|^2-|h_d^0|^2)^2 +\Delta V\;, 
\eea 
where the radiative corrections (in the one-loop effective potential
approximation) are given in the $\overline{DR}$ scheme by, 
\be 
\Delta V=\sum_{i}\frac{(-1)^{2s_i}}{64\pi^2}
Tr\left( (\CM_i\CM_i^\dagger )^2
\left[\log \frac{\CM_i\CM_i^\dagger}{Q^2}-\frac{3}{2}\right]\right)\;.  
\ee
Here, the sum over $i$ runs over all fields that couple to Higgs fields,
$\CM_i\CM_i^\dagger$ is the {\it Higgs field dependent} mass squared
matrix (defined as the second derivative of the tree level potential),
and the trace is over the internal as well as any spin indices.  One may
compute the gauge boson masses in terms of the Higgs field vacuum
expectation values $v_u$ and $v_d$ by minimizing the scalar potential in
the $h_u^0$ and $h_d^0$ directions.  This leads to the well-known
condition 
\be 
\frac{M_Z^2}{2} =
\frac{(m_{H_d}^2+\Sigma_d^d)-(m_{H_u}^2+\Sigma_u^u)\tan^2\beta}{\tan^2\beta
-1} -\mu^2 \; .
\label{eq:mssmmu}
\ee 
Here the $\Sigma_u^u$ and $\Sigma_d^d$ terms arise from first derivatives
of $\Delta V$ evaluated at the potential minimum and $\tan\beta\equiv
v_u/v_d$.  At the one-loop level, $\Sigma_u^u$ contains the
contributions $\Sigma_u^u(\tst_{1,2})$, $\Sigma_u^u(\tb_{1,2})$,
$\Sigma_u^u(\ttau_{1,2})$, $\Sigma_u^u(\tw_{1,2})$,
$\Sigma_u^u(\tz_{1-4})$, $\Sigma_u^u(h,H)$, $\Sigma_u^u(H^\pm)$,
$\Sigma_u^u(W^\pm)$, $\Sigma_u^u(Z)$, and $\Sigma_u^u(t)$. $\Sigma_d^d$
contains similar terms along with $\Sigma_d^d(b)$ and $\Sigma_d^d(\tau)$
while $\Sigma_d^d(t)=0$~\cite{bbhmt}. 

Although we have highlighted third generation matter sfermion
contributions here because these frequently dominate on account of their
large Yukawa couplings, we note that there are also first/second
generation contributions $\Sigma_u^u(\tq,\tell)$ and $\Sigma_d^d(\tq
,\tell )$ that arise from the quartic $D$-term interactions between the
Higgs sector and matter scalar sector even when the corresponding Yukawa
couplings are negligibly small. These contributions are proportional to
$(T_{3_i} - Q_i\sin^2\theta_W)\times F(m_i^2)$, 
where $T_{3_i}$ is the hypercharge, $Q_i$ is the electric charge and 
$F(m^2)=m^2(\log\frac{m^2}{Q^2}-1)$ 
of the $i^{th}$ matter scalar. 
%and $i$ runs over the matter scalars. 
Although the
scale of these is set by the electroweak gauge couplings rather than the
top Yukawa coupling, these can nevertheless be sizeable if the squarks
of the first two generations are significantly heavier than third
generation squarks. However, in models such as mSUGRA -- where all squarks
of the first two generations (and separately, the corresponding sleptons)
are nearly mass degenerate -- these contributions largely cancel.  Indeed, the
near cancellation (which would be perfect cancellation in the case of
exact degeneracy) occurs within each generation, and separately for
squarks and for sleptons. These terms, summed over each of the first two
generations, are always smaller than the other terms in the $C_i$
and $B_i$ arrays used to define our fine-tuning criterion below, 
and so do not alter our fine-tuning measure defined below.

The reader may wonder that we are treating the first two generations
differently from the third generation in that for the latter we consider
the contributions from each squark separately ({\it i.e.} not allow for
cancellations of the contributions to say $\Sigma_u^u$ from different
squarks to cancel), while we sum the contributions from the entire
first/second generation to obtain a tiny contribution. The reason for
this is that the mSUGRA framework {\em predicts} degenerate first/second
generation squarks (and sleptons) while the top squark masses (remember
that top squarks frequently make the largest contribution to
$\Sigma_u^u$) are essentially independent. In an unconstrained framework
such as the pMSSM~\cite{pmssm} we {\it would not combine} the
contributions from the first/second generation scalars; if these are
very heavy and have large intra-generation splitting, their contribution
to $\Delta_{\rm EW}$ can be significant.

\subsection{Electroweak scale fine-tuning}
\label{ssec:ewft}

One measure of fine-tuning, introduced previously in
Ref.~\cite{bbhmt,kn}, is to posit that there are no large cancellations
in Eq.~(\ref{eq:mssmmu}). This implies that all terms on the right-hand side
to be comparable to $M_Z^2/2$, {\it i.e.} that each of the three tree level
terms $C_{H_d}\equiv |m_{H_d}^2/(\tan^2\beta -1)|$, $C_{H_u}\equiv
|-m_{H_u}^2\tan^2\beta /(\tan^2\beta -1)|$, $C_\mu\equiv |-\mu^2 |$ and
each $C_{\Sigma_{u,d}^{u,d} (i)}$ is less
than some characteristic value $\Lambda$ where $\Lambda\sim M_Z^2$. 
(Here, $i$ labels SM and supersymmetric
particles that contribute to the one-loop Higgs potential and includes 
the sum over matter sfermions from the first two generations.)
This leads to a fine-tuning measure 
\be \Delta_{\rm EW}\equiv max(C_i )/(M_Z^2/2).  
\label{eq:ewft}
\ee 
A feature of defining the fine-tuning parameter solely in terms of weak
scale parameters is that it is independent of whether the SUSY particle
spectrum is generated using some high scale theory or generated at or
near the weak scale, as in the pMSSM or possibly in gauge-mediation~\cite{gmsb}: if
the spectra and weak scale couplings from two different high scale
theories are identical, the corresponding fine-tuning measures are the same. However, as
we will see in Subsection~\ref{ssec:hsft}, in theories such as mSUGRA
$\Delta_{\rm EW}$ does not capture the entire fine-tuning because
Eq.~(\ref{eq:mssmmu}) does not include information about the underlying
origin of the weak scale mass parameters.

It is worthwhile to note that over most of parameter space
the dominant contribution to $\Delta_{\rm EW}$
comes from the weak scale values of $m_{H_u}^2$ and $\mu^2$. To see
this, we note that unless 
$\tan\beta$ is very small, aside from radiative
corrections, we would have simply that $M_Z^2/2\simeq -m_{H_u}^2-\mu^2$.
As is customary, the
value of $\mu^2$ is selected so that the correct value of $M_Z$ is
generated.  In this case, over much of parameter space $\Delta_{\rm EW}\sim
|\mu^2|/(M_Z^2/2)$.  Only when $|\mu |$ becomes small do the radiative
corrections become important -- providing the largest contribution to
Eq.~(\ref{eq:mssmmu}). Thus, contours of fixed
$\Delta_{\rm EW}$ typically track the contours
of $|\mu|$ except when $|\mu|$ is small; in this latter case, $\Delta_{\rm EW}$ is
determined by the $|\Sigma_u^u|$ whose value is loop-suppressed.
In Fig.~\ref{fig:mu} we show the surface of
$|\mu|$ values in the $m_0\ vs.\ m_{1/2}$ plane of mSUGRA/CMSSM for
$A_0=0$ and $\tan\beta =10$.  Here, $\mu$ is small either at low $m_0$
and $m_{1/2}$ (the bulk region\cite{bulk}), or  in the HB/FP 
region\cite{hb_fp} at large values of $m_0$.
\FIGURE[tbh]{
\includegraphics[width=12cm,clip]{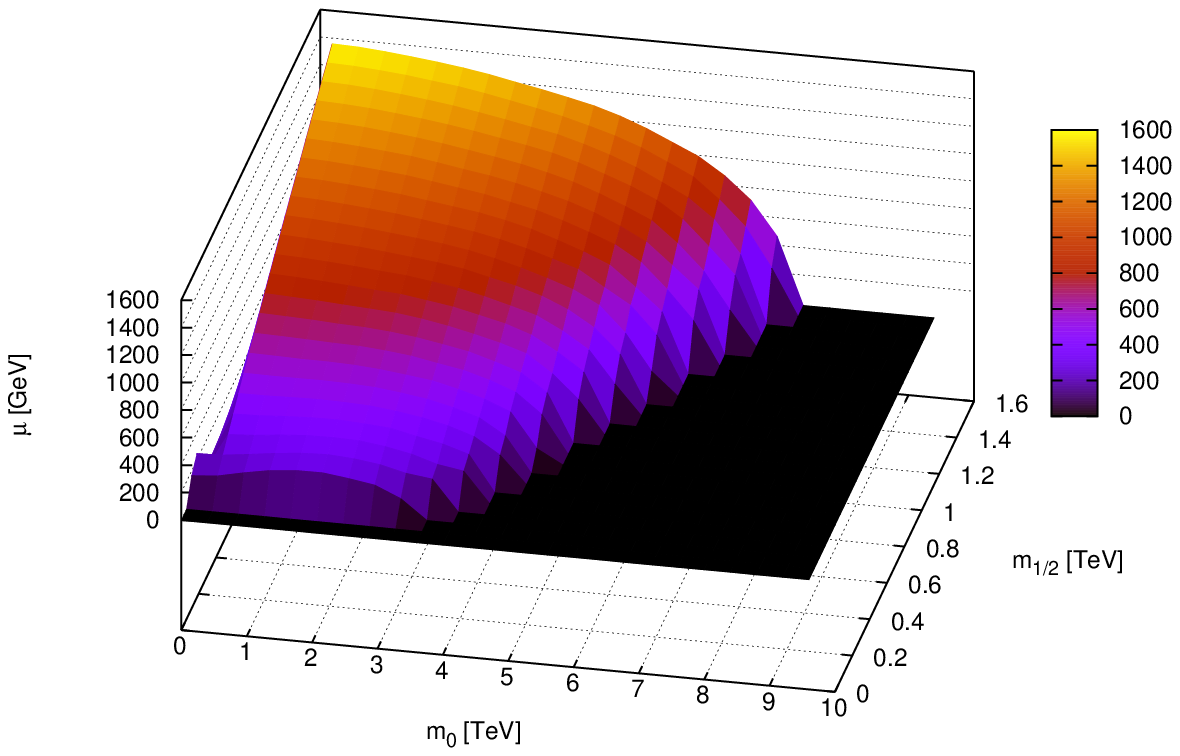}
\caption{The value of $\mu$ in the $m_0\ vs.\ m_{1/2}$ plane
of mSUGRA for $A_0=0$ and $\tan\beta =10$. 
We set $\mu=0$ in theoretically forbidden regions.}
\label{fig:mu}}

\subsection{High scale fine-tuning}
\label{ssec:hsft}

As mentioned above, Eq.~(\ref{eq:mssmmu}) is obtained from the weak
scale MSSM potential and so contains no information about its possible
high scale origin. To access this, and make explicit the dependence on
the high scale $\Lambda$, we must write the {\it weak scale} parameters
$m_{H_{u,d}}^2$ in Eq.~(\ref{eq:mssmmu}) as $$m_{H_{u,d}}^2=
m_{H_{u,d}}^2(\Lambda) +\delta m_{H_{u,d}}^2, \
\mu^2=\mu^2(\Lambda)+\delta\mu^2\;,$$ where
$m_{H_{u,d}}^2(\Lambda)$ and $\mu^2(\Lambda)$ are the corresponding
parameters renormalized at the high scale $\Lambda$. It is the $\delta
m_{H_{u,d}}^2$ terms that contain the $\log\Lambda$ dependence shown in
the $C_2$ type terms in Eq.~(\ref{eq:generic}). In this way, we get
\be 
\frac{M_Z^2}{2} = \frac{(m_{H_d}^2(\Lambda)+ \delta m_{H_d}^2 +
\Sigma_d^d)-(m_{H_u}^2(\Lambda)+\delta m_{H_u}^2+\Sigma_u^u)\tan^2\beta}{\tan^2\beta -1} 
-(\mu^2(\Lambda)+\delta\mu^2)\;.
\label{eq:FT}
\ee 
Following the same spirit that we had used in our earlier analyses~\cite{bbhmt}, 
we can now define a fine-tuning measure that encodes the
information about the high scale origin of the parameters by requiring
that each of the terms on the right-hand-side of Eq.~(\ref{eq:FT}) to be
smaller than a pre-assigned $\Delta_{\rm HS}$ times $\frac{M_Z^2}{2}$. The
high scale fine-tuning measure $\Delta_{\rm HS}$ is thus defined to be
\be 
\Delta_{\rm HS}\equiv max(B_i )/(M_Z^2/2)\;, 
\label{eq:hsft} 
\ee 
with 
\begin{eqnarray*}
B_{H_d}\equiv|m_{H_d}^2(\Lambda)/(\tan^2\beta -1)|,& &B_{\delta H_d}\equiv |\delta
m_{H_d}^2/(\tan^2\beta -1)|,\\
B_{H_u}\equiv|-m_{H_u}^2(\Lambda)\tan^2\beta /(\tan^2\beta -1)|,& &B_{\delta H_u}\equiv
|-\delta m_{H_u}^2\tan^2\beta /(\tan^2\beta -1)|,\ etc.,
\end{eqnarray*}
defined analogously
to the set  $C_{i}$ in Sec.~\ref{ssec:ewft}. As discussed above, in
models such as mSUGRA whose domain of validity extends to very high scales,
because of the large logarithms one would expect that (barring seemingly
accidental cancellations) the $B_{\delta H_u}$ 
contributions to $\Delta_{\rm HS}$ would be much larger than any
contributions to $\Delta_{\rm EW}$ because the $m_{H_u}^2$ evolves from
$m_0^2$ to negative values. 

As we have noted, $\Delta_{\rm EW}$ indeed provides a measure of EWFT
that is determined only by the sparticle spectrum: by construction, it
has no information about any tuning that may be necessary in order to
generate a given weak scale SUSY mass spectrum. Thus, while 
{\em for a given SUSY spectrum} $\Delta_{\rm EW}$ includes information about the 
{\it minimal amount of fine-tuning} that is present in the model,
$\Delta_{\rm HS}$ better represents the fine-tuning that is present in
high scale models.

The reader may have noticed that -- unlike in our definition of
$\Delta_{\rm EW}$ in Eq.~(\ref{eq:ewft}) where we have separated out
the contributions from various sources and required each of these to not
exceed some preassigned value -- we have neglected to separate out the
various contributions to $\delta m_{H_{u,d}}^2$ that determine
$\Delta_{\rm HS}$. We have done so mainly for
convenience,\footnote{Unlike for $\Delta_{\rm EW}$ where we have
separated the contributions by particles (and treated these as
independent) for the electroweak scale theory, in a constrained high
scale model, these would not be independent. Instead, we could separate
out contributions that have independent origins in the high scale
model. For instance, for the mSUGRA model we should separately require contributions 
from gauginos, scalars and $A$-parameters to $\delta m_{H_i}^2$ to be
small. We have not done so here mainly for expediency. In this sense if
accidental cancellations reduce $\Delta_{\rm HS}$ to very small values,
this should be interpreted with care.} but this will also help us to connect
up with what has been done in the literature.

Before closing this section, we remark that our definition of
$\Delta_{\rm HS}$ differs in spirit from that used by some groups\cite{many}. 
These authors write the $m_{H_u}^2$ as a quadratic
function of the high scale parameters $\xi_i = \{m_0, \ m_{1/2}, \ A_0\}$
  for mSUGRA, {\it i.e.} 
\be 
m_{H_u}^2 =  \sum a_{ij} \xi_i\xi_j\;, 
\label{eq:otherft}
\ee 
and substitute this (along with the corresponding form for $m_{H_d}^2$) 
in Eq.~(\ref{eq:mssmmu}) to examine the sensitivity of
$M_Z^2$ to changes in the high scale parameters.\footnote{Typically
these authors use $\Delta \equiv \frac{a_i}{M_Z^2}\frac{\partial
M_Z^2}{\partial a_i}$ (where $a_i$ labels the input parameters) 
as a measure of the sensitivity to parameters\cite{BG}. 
This prescription agrees with our $\Delta$ at tree level, but differs when
loop corrections are included.} In the resulting expression, the
coefficient of $m_0^2$ in Eq.~(\ref{eq:otherft}) is often very small
because of cancellations with the large logarithms, suggesting that the region
of mSUGRA with rather large $m_0$ (but small $m_{1/2}$ and $A_0$) is not
fine-tuned: we feel that this is misleading and so have {\it separated}
the contributions from the large logarithms in our definition of
$\Delta_{\rm HS}$.  Combining all $m_0^2$ contributions into a single term
effectively combines $m_i^2(\Lambda) +\delta m_i^2$ into a single
quantity which (aside from the one-loop terms $\Sigma_u^u $ and
$\Sigma_d^d$) evidently is {\it the weak scale value of $m_i^2$} in our
definition of $\Delta_{\rm HS}$. Except for these one-loop correction
terms, $\Delta_{\rm HS}$ then reduces to $\Delta_{\rm EW}$!

In defining $\Delta_{\rm HS}$ as above, we have taken the view that
the high scale parameters as well as the scale at which we assume the
effective theory to be valid are independent. {\it In the absence of an
underlying theory of the origin of these parameters}, we regard
cancellations between terms in Eq.~(\ref{eq:otherft}) that occur for
{\it ad hoc} relations\footnote{It may be argued that such an analysis
is helpful as a guide to model builders attempting to construct models
of natural SUSY.} between model parameters and lead one to conclude that
$M_Z$ is not fine-tuned as fortuitous, and do not incorporate it into
our definition of high scale fine-tuning. We emphasize that we would
view the fine-tuning question very differently if indeed the high scale
parameters were all related from an underlying 
meta-theory.\footnote{This situation seems to occur in the so-called
mixed-modulus-anomaly mediated SUSY breaking models 
for some ranges of the mixing parameter $\alpha$ as emphasized in
Ref.~\cite{nilles}.} In that case, though, as we just mentioned,
$\Delta_{\rm EW}$ would be an adequate measure of  fine
tuning.

%%%%%%%%%%%%%%%%%%%%%%%%%%%%%%%%%%%%%%%%%%%%%%%%%
\section{Results in $m_0\ vs.\ m_{1/2}$ plane}
\label{sec:plane}
%%%%%%%%%%%%%%%%%%%%%%%%%%%%%%%%%%%%%%%%%%%%%%%%%

We present our first results as contours of $\Delta_{\rm HS}$ and
$\Delta_{\rm EW}$ in the $m_0\ vs.\ m_{1/2}$ plane of the mSUGRA/CMSSM
model. For all plots, we take $m_t=173.2$~GeV and we generate SUSY
particle mass spectra from Isasugra v7.83~\cite{isasugra}.  In
Fig.~\ref{fig:tanb10_A0}, we show contours of $\Delta_{\rm HS}$ in frame
{\it a}) and for $\Delta_{\rm EW}$ in frame~{\it b}). For both frames,
we take $A_0=0$, $\tan\beta =10$ and $\mu >0$. The gray-shaded regions
running from the extreme left of the plot, across the bottom and on to
the right are excluded by either a $\ttau_1$ as LSP (left-side), LEP1
constraints (bottom) or lack of appropriate EWSB (right-side). The
region marked LEP2 is excluded by LEP2 chargino seaches
($m_{\tw_1}>103.5$~GeV)~\cite{lep2}. The region below the contour
labeled LHC7 is excluded by lack of a SUSY signal from SUSY searches at
LHC7 with 5~fb$^{-1}$ of data\cite{atlas_susy,cms_susy}. The dashed
portion of the contour is our extrapolation of LHC7 results to higher
values of $m_0$ than are shown by the Atlas/CMS collaborations. 
We also denote regions where the calculated\cite{tata} branching fraction $B_s\to\mu^+\mu^-$ falls outside
its newly measured range from LHCb observations\cite{lhcb}, which now require
\be
2\times 10^{-9}<BF(B_s\to\mu^+\mu^- )<4.7\times 10^{-9}\ \ \ (95\%\ CL) .
\ee
However, for the low value of $\tan\beta$ in this figure 
(and also in subsequent figures with $\tan\beta=10$) the LHCb does not lead to any constraint 
because the SUSY contribution, which grows rapidly with $\tan\beta$, is rather small. 
The green-shaded region is where the thermally-generated relic density of
neutralinos (computed using IsaReD\cite{isared}) satisfies
$\Omega^{th}_{\tz_1}h^2 < 0.1194$, the $2\sigma$ upper limit on the density
of cold dark matter obtained by the WMAP collaboration~\cite{wmap}.
This region encompasses the stau-coannihilation strip~\cite{stau}
(extreme left), the bulk region~\cite{bulk} (bottom left corner) and the
well-known focus point/hyperbolic branch region~\cite{hb_fp} of the
model.  The shaded region labeled $a_\mu$ is where the measured muon
magnetic moment~\cite{bnl} satisfies 4.7$\times 10^{-10} \leq a_\mu \leq
52.7\times 10^{-10}$, within $3\sigma$ of its theoretical
value~\cite{davier}. For $A_0=0$ adopted in this figure, $m_h < 123$~GeV
over the entire parameter plane,
so that mSUGRA is excluded for $A_0\sim 0$ (as
noted in Ref.~\cite{bbm}) unless one has very high values of
$m_0$ and $m_{1/2}$~\cite{eo}. 

As might be anticipated, $\Delta_{\rm HS}$ grows with increasing values
of $m_0$ or $m_{1/2}$, so that we expect contours of fixed $\Delta_{\rm
HS}$ to be oval-shaped in the $m_0-m_{1/2}$ plane. This is readily seen
in frame {\it a}) of Fig.~\ref{fig:tanb10_A0}, except that because the
oval is extremely elongated since the scales on the two axes are very
different, we see only a small  part of this contour (which appears as
nearly vertical lines) for very large values of $\Delta_{\rm HS}$.   
We have checked that 
$\Delta_{\rm HS} < 150$ is already excluded by LHC searches, so high
scale fine-tuning of less than a percent is now mandatory for
$A_0=0$. If we take the high scale origin of the mSUGRA model seriously,
we see that without a theory that posits special relations between the
parameters that could lead to automatic cancellation of the large
logarithms that enter $\Delta_{\rm HS}$, we are forced to conclude that
LHC data imply that the theory is fine-tuned to a fraction of a percent.
For the portion of the plane compatible with LHC constraints on
sparticles, the smallest values of $\Delta_{\rm HS}$ occur where $m_0$
and $m_{1/2}$ are simultaneously small. As $m_0$ moves to the multi-TeV
scale, $\Delta_{\rm HS}$ exceeds 1000, and fine-tuning of more than part
per mille is required.

In frame {\it b}) of the figure, we show contours of constant
$\Delta_{\rm EW}$.  Over most of the plane, these contours tend to track
contours of constant $\mu^2$ since $M_Z^2/2\sim -m_{H_u}^2-\mu^2$ so
that when $|m_{H_u}^2|\gg M_Z^2/2$, then $-m_{H_u}^2\sim \mu^2$.  Thus,
along the contours of $\Delta_{\rm EW}$, the value of $m_{H_u}^2$ is
independent of $m_0$ at least until the contours turn around at large
values of $m_0$ and $m_{1/2}$. This is just the focus point behaviour
discussed in the second paper of Ref.~\cite{hb_fp}.\footnote{More precisely, the discussion in
this paper was for a fixed value of $m_{1/2}$ so that the range of $m_0$
was limited because we hit the theoretically excluded region. We see
though that the same value of $m_{H_u}^2$ can be obtained if we
simultaneously increase $m_0$ and $m_{1/2}$ so that we remain in the
theoretically allowed region.} The $\Delta_{\rm EW}$ contours, for large
values of $m_0$ bend over and track excluded region on the right where
$\mu^2$ becomes negative. This is the celebrated hyperbolic branch~\cite{hb_fp} of small $|\mu|$. 
The contours of $\Delta_{\rm EW}$ then bend around for very large values
of $m_0$ because $\Sigma_u^u$ contributions, especially from $\tst_2$
loops --- increase with $m_0$ --- begin to exceed $-m_{H_u}^2 \simeq
\mu^2$.  Indeed, Fig.~\ref{fig:tanb10_A0}{\it b}) shows that there is a
region close to (but somewhat removed from) the ``no EWSB'' region
on the right where $\Delta_{\rm EW}$ becomes anomalously small even for
large values of $m_0$ and $m_{1/2}$. 
It is instructive to see that while this low EWFT region is close to the
relic-density consistent region with small $\mu$~\cite{hb_fp}, it is
still separated from it.\footnote{Much of the literature treats these
regions as one. While this is fine for some purposes, it seems necessary
to be clear on the difference when discussing either dark matter or
EWFT. Note that $\Delta_{\rm HS}$ is large in both regions.}  While
$\Delta_{\rm EW}\sim 100$ is excluded at low $m_0$, this 1\% EWFT
contour, even with the resolution of our scan, extends out to very large
$m_0\sim 6$~TeV values for $m_{1/2}$ as high as 1~TeV! While these plots
show that relatively low EWFT ($\Delta_{\rm EW}$ of a few tens) is still
allowed by LHC7 constraints on sparticles, it is important to realize
that {\it these planes are now excluded since they cannot accommodate
$m_h\sim 125$~GeV}.
\FIGURE[tbh]{
\includegraphics[width=7cm]{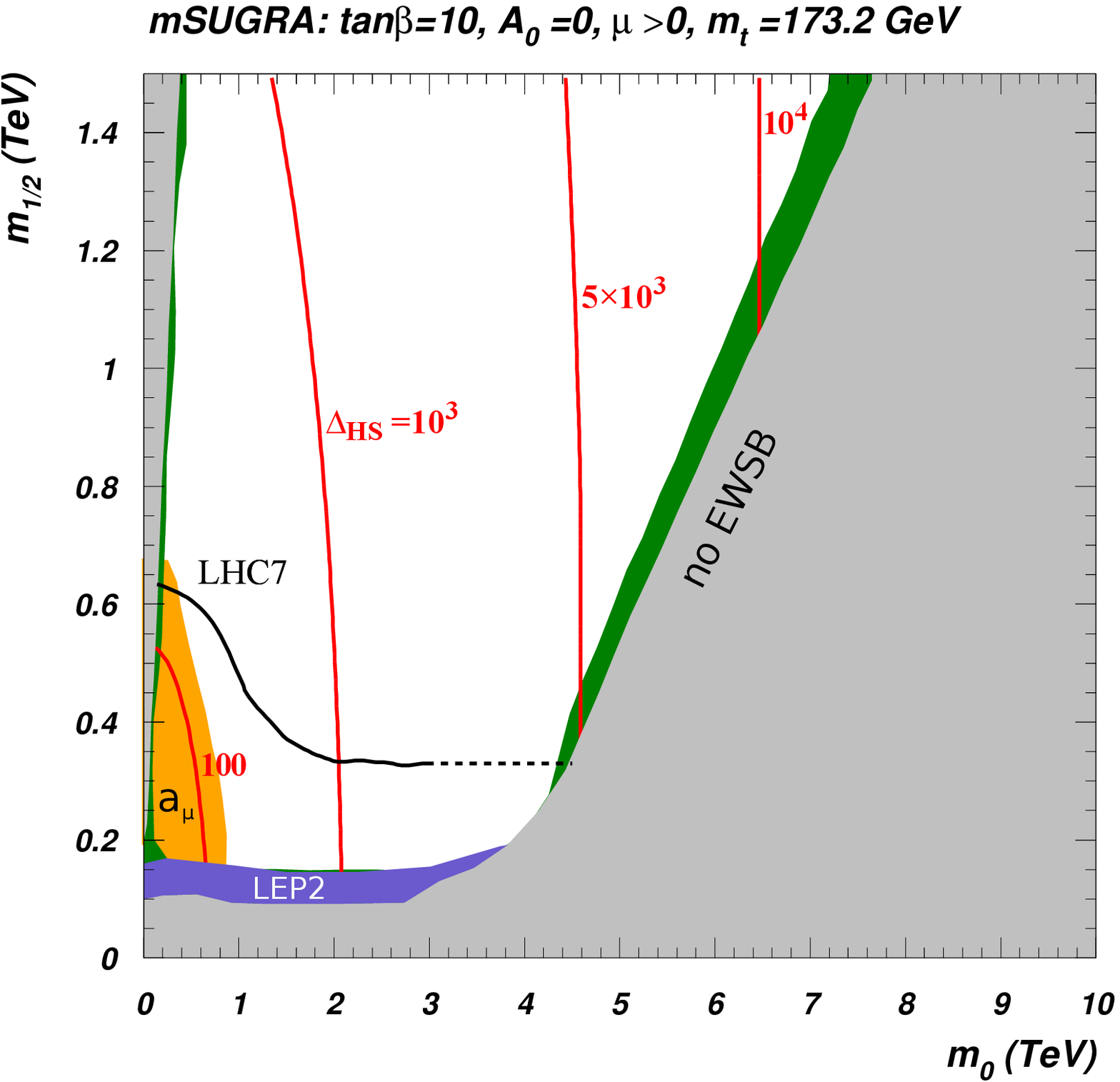}
\includegraphics[width=7cm]{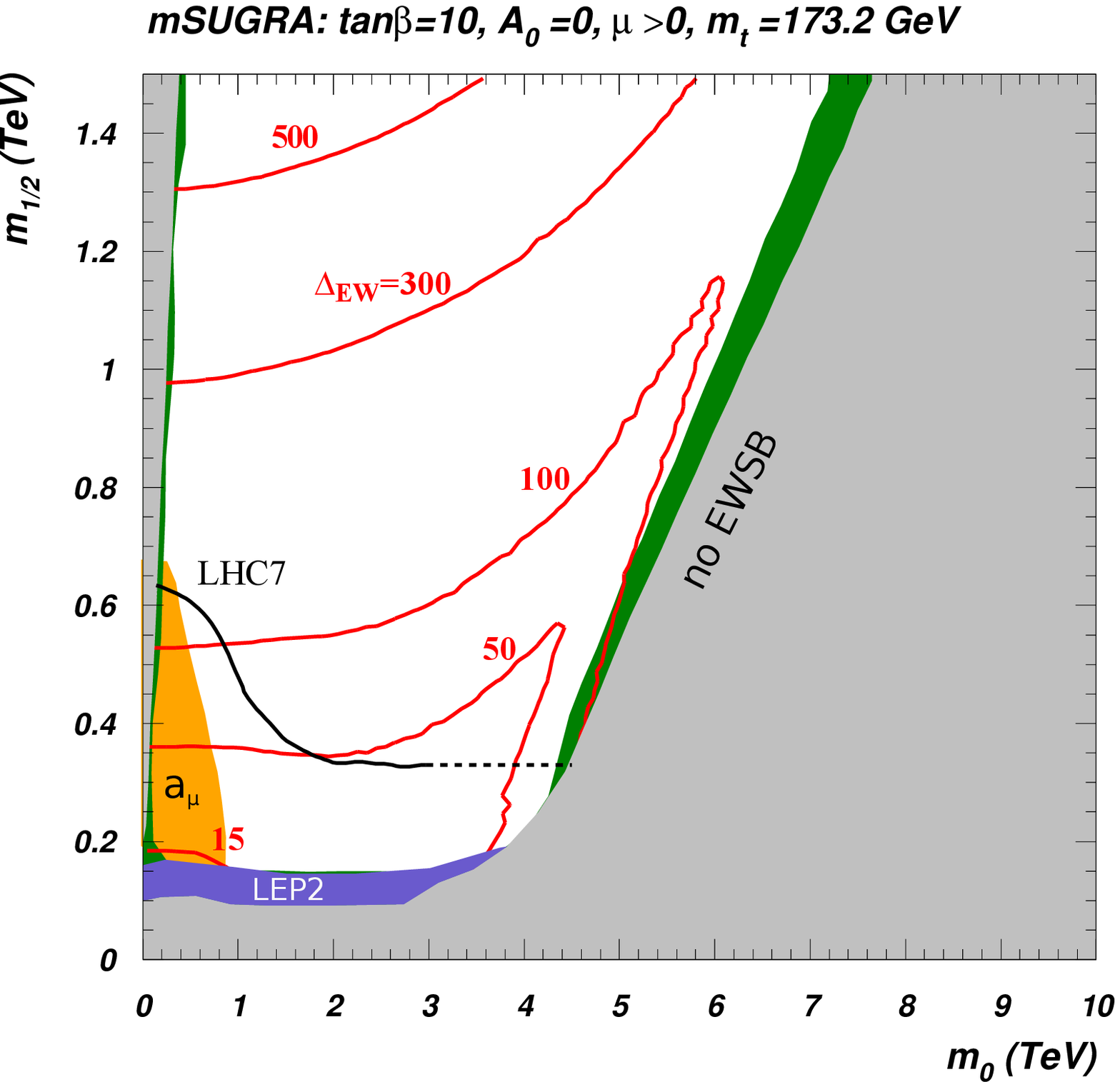}
\caption{Contours of {\it a}) $\Delta_{\rm HS}$ and {\it b}) $\Delta_{\rm EW}$
in the mSUGRA model with $A_0=0$ and $\tan\beta =10$.  We take $\mu >0$
and $m_t=173.2$~GeV. The grey region on the left is excluded either
because the stau is too light or becomes tachyonic, the grey region at
the bottom is excluded  by LEP1 constraints, while in the region
on the right we do not get the correct pattern of EWSB, since either
$\mu^2$ or $m_A^2$ become negative. The region labeled LEP2 is excluded
by constraints on the chargino mass. The region labeled $a_\mu$ is
allowed at the 3$\sigma$ level by the E821 experiment while in the
green-shaded region, the thermal neutralino relic density is
at or below the WMAP measurement of the cold dark matter density. 
The region below black contour labeled LHC7 is excluded by SUSY searches. 
The lighter Higgs boson mass $m_h< 123$~GeV throughout 
this parameter plane.}
\label{fig:tanb10_A0}}

Before moving on to other planes, we remark that for the smallest values of
$m_0$ in the LHC-allowed regions of the figure, $\Delta_{\rm HS} \sim
\Delta_{\rm EW}$. As we have explained, $\Delta_{\rm HS}$ is determined
by the value of $|\delta m_{H_u}^2|$ (see Eq.~\ref{eq:FT}), which for $m_0
\sim 0$ is just $|m_{H_u}^2|$ that determined $\Delta_{\rm EW}$ when $m_0$
is very small. We thus see that the two measures are roughly comparable
for small values of $m_0$ but deviate from one another as $m_0$ is
increased. We see that $\Delta_{\rm HS}$
typically exceeds $\Delta_{\rm EW}$ by an order of magnitude, because 
of the large logarithm of the ratio of the GUT and weak scales, 
except in the HB/FP region where $\Delta_{\rm EW}$ is exceptionally small.

In Fig.~\ref{fig:tanb50_A0} we show the $m_0\ vs.\ m_{1/2}$ plane for
$\tan\beta =50$ and $A_0=0$. The contours in both frames are
qualitatively very similar those for the $\tan\beta =10$ case. As
expected, regions of low $\Delta_{\rm EW}$ extend to very large $m_0$ and
$m_{1/2}$ in the HB region.  One difference from the $\tan\beta=10$ case
discussed above is that this time the HB region largely overlaps with the
relic-density-consistent green-shaded region. 
Note also that for this large value of $\tan\beta$ there is
a considerable region (left of the LHCb contour) that 
is now excluded due to too large a value of $BF(B_s\to\mu^+\mu^-)$. Again, the entire
region of plane shown is excluded by the LHC Higgs discovery at 125~GeV.
\FIGURE[tbh]{
\includegraphics[width=7cm]{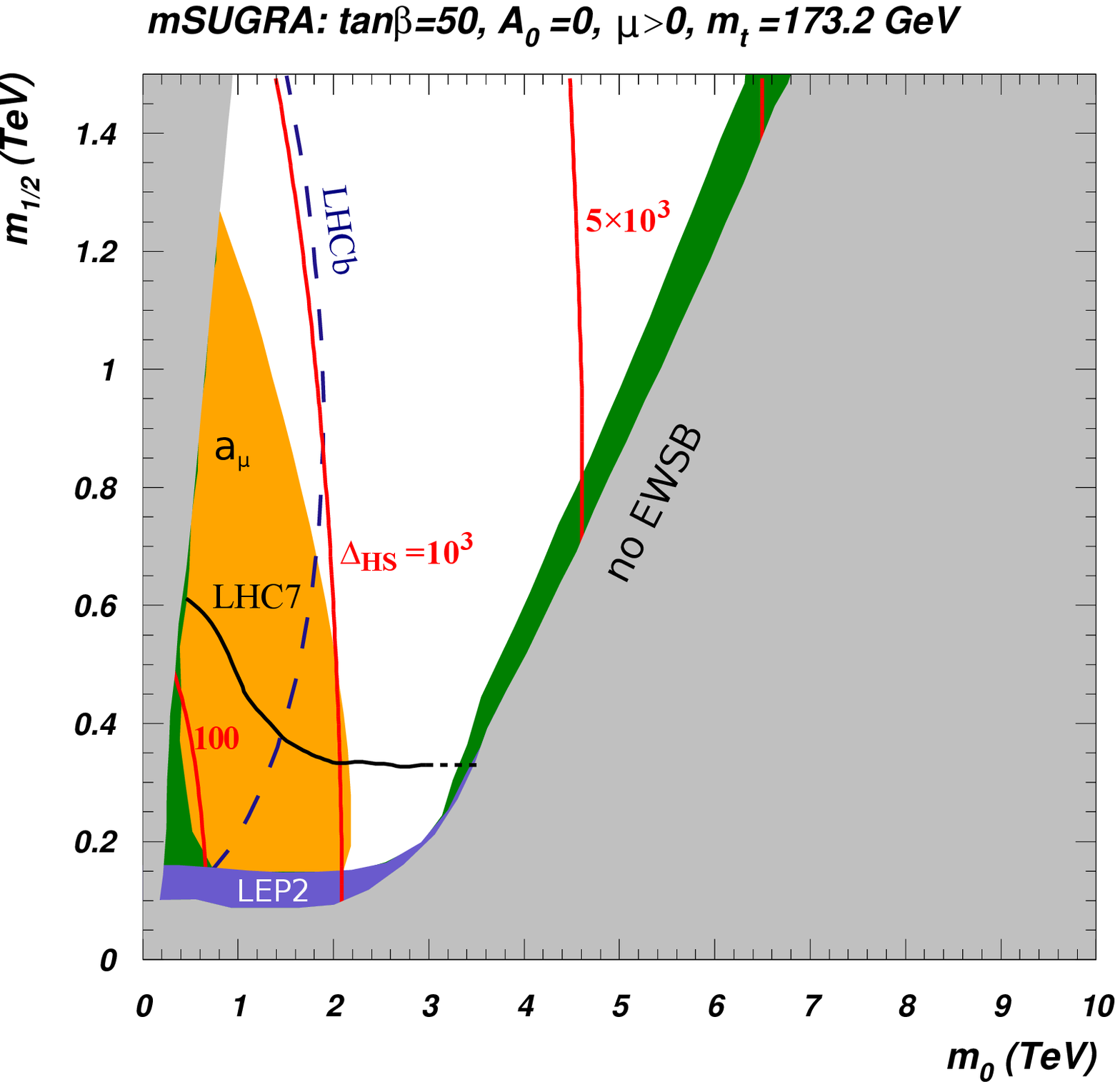}
\includegraphics[width=7cm]{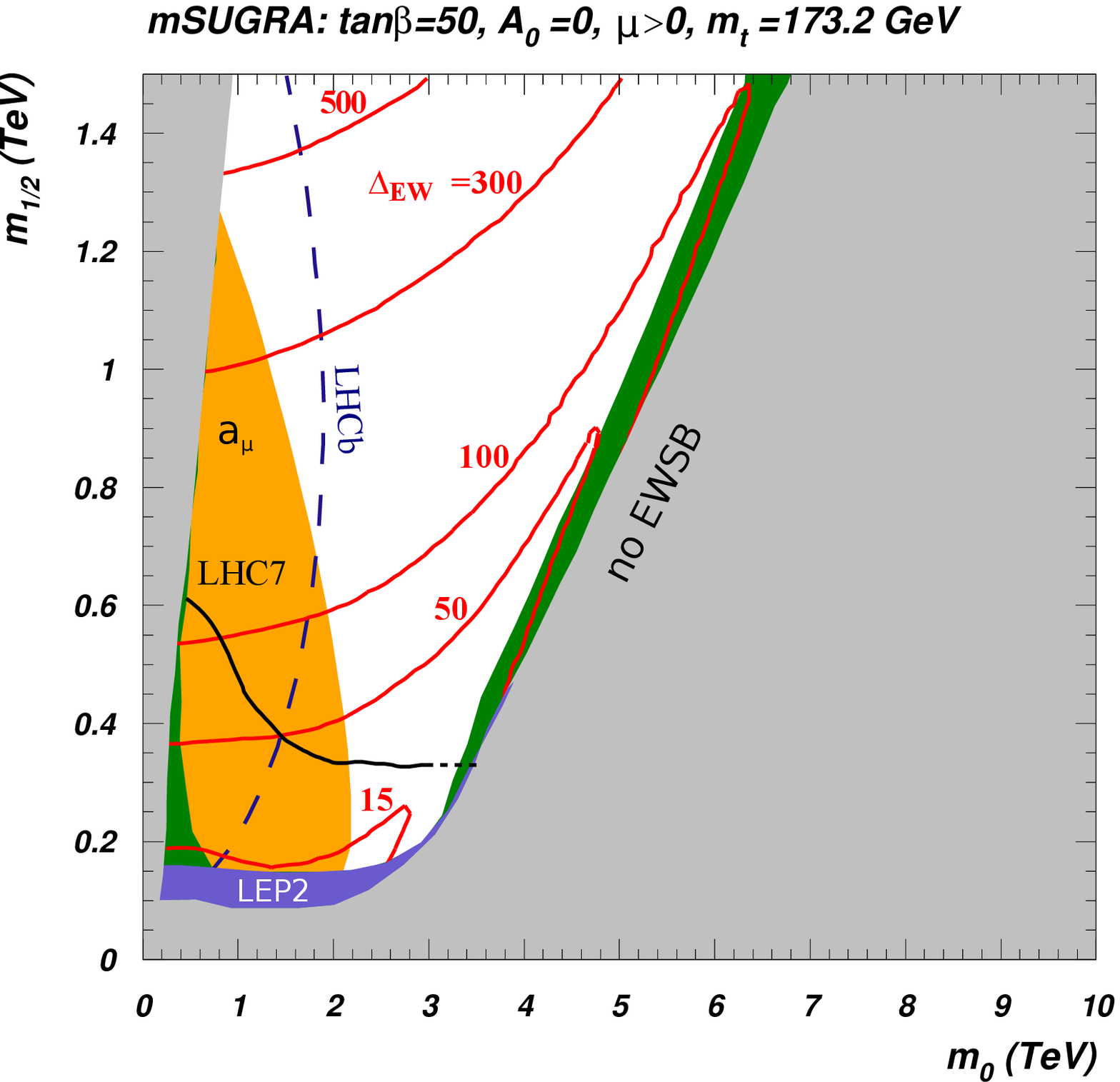}
\caption{Contours of {\it a}) $\Delta_{\rm HS}$ and {\it b}) $\Delta_{\rm EW}$ 
in the mSUGRA model with $A_0=0$, $\tan\beta =50$ and $\mu>0$.
The value of $m_t$  as well as the various shaded/coloured regions are
as in Fig.~\ref{fig:tanb10_A0}. The region to the left of the long dashed blue contour is
excluded by LHCb measurements.}
\label{fig:tanb50_A0}}

In Fig.~\ref{fig:tanb10_Am1}, we show contours of $\Delta_{\rm HS}$ and
$\Delta_{\rm EW}$ for $\tan\beta =10$ and $A_0=-m_0$.  The first thing
to notice is that the HB/FP region does not appear. The region at
extremely large $m_0$ is still theoretically excluded, but more
typically because $m_A^2$ turns negative (or there are tachyons) 
{\em not because $\mu^2$ turns negative}.\footnote{For $m_{1/2}=500$~GeV, this
happens for $m_0\agt 22$~TeV. We mention that this breakdown of
parameter space could be an artifact of the ISAJET algorithm for 
computing the sparticle mass spectrum in mSUGRA. 
An approximate tree-level spectrum is first required in order to evaluate
the radiative corrections that can potentially yield a valid solution
using an iterative procedure. But in the absence of a non-tachyonic,
tree-level spectrum with the correct EWSB pattern, the program is unable
to compute the radiatively corrected mass spectrum.} 
In addition,
the very large $m_0\agt 7-9$~TeV region yields a value of $m_h>123$~GeV:
thus, the bulk of this plane is still excluded.  The contours of
$\Delta_{\rm HS}$ are qualitatively similar to the $A_0=0$ cases, and LHC7
still excludes $\Delta_{\rm HS}<100$, so again a HSFT of more than 1\% is
required.  In the region with $m_h>123$~GeV, $\Delta_{\rm HS}\agt 1.5\times
10^4$, and extreme HSFT is required. Moving to frame {\it b}), we note
that though the contours of fixed $\Delta_{\rm EW}$ now run from top
left to lower right, these still follow the lines of fixed values of
$\mu^2$.  Moreover, values of $\Delta_{\rm EW}$ below 100 are 
excluded by just the LHC7 sparticle mass constraints. 
If one also imposes $m_h>123$~GeV, then
$\Delta_{\rm EW}\agt 2000$ is required over the entire plane shown.
\FIGURE[tbh]{
\includegraphics[width=7cm]{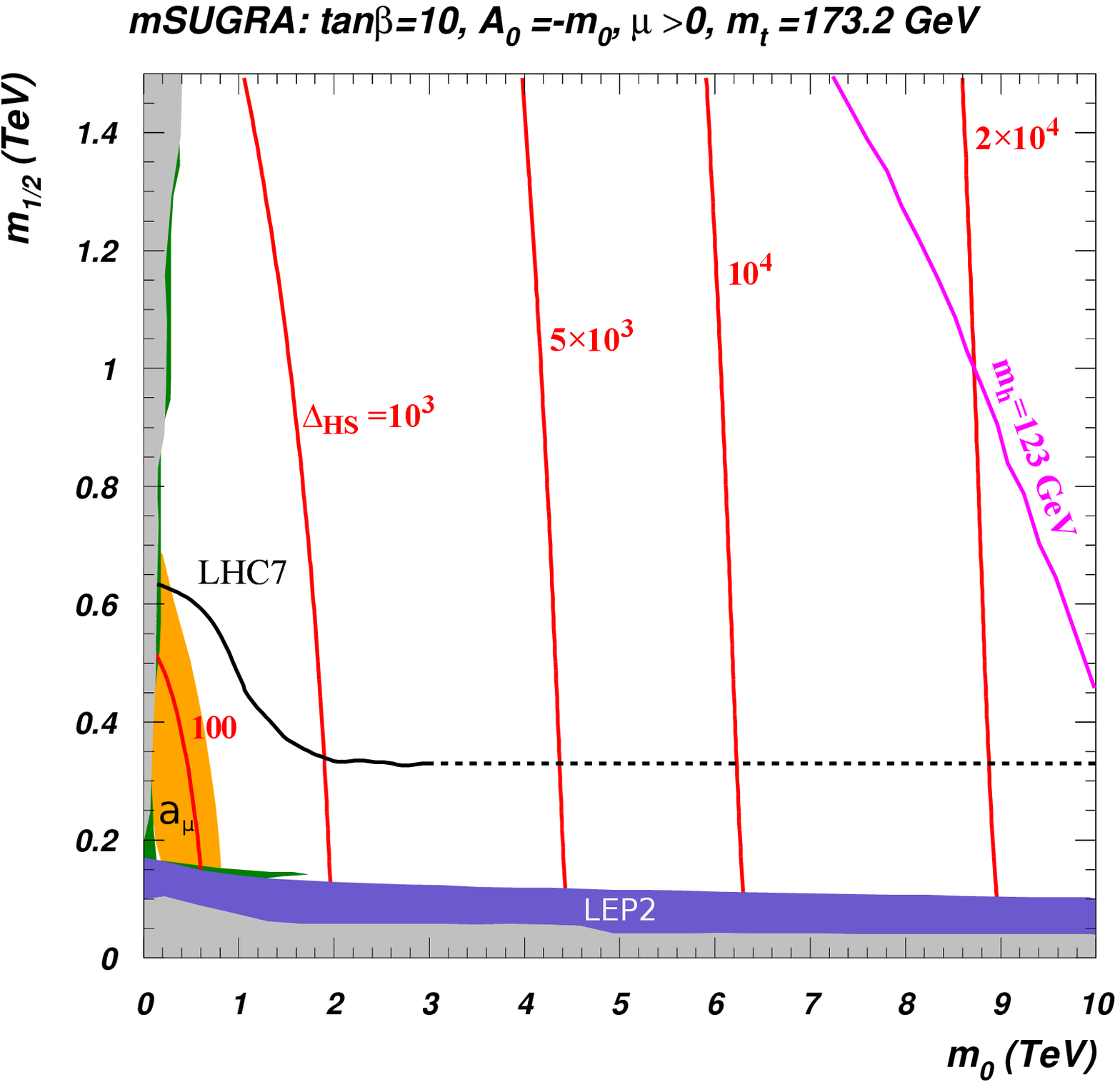}
\includegraphics[width=7cm]{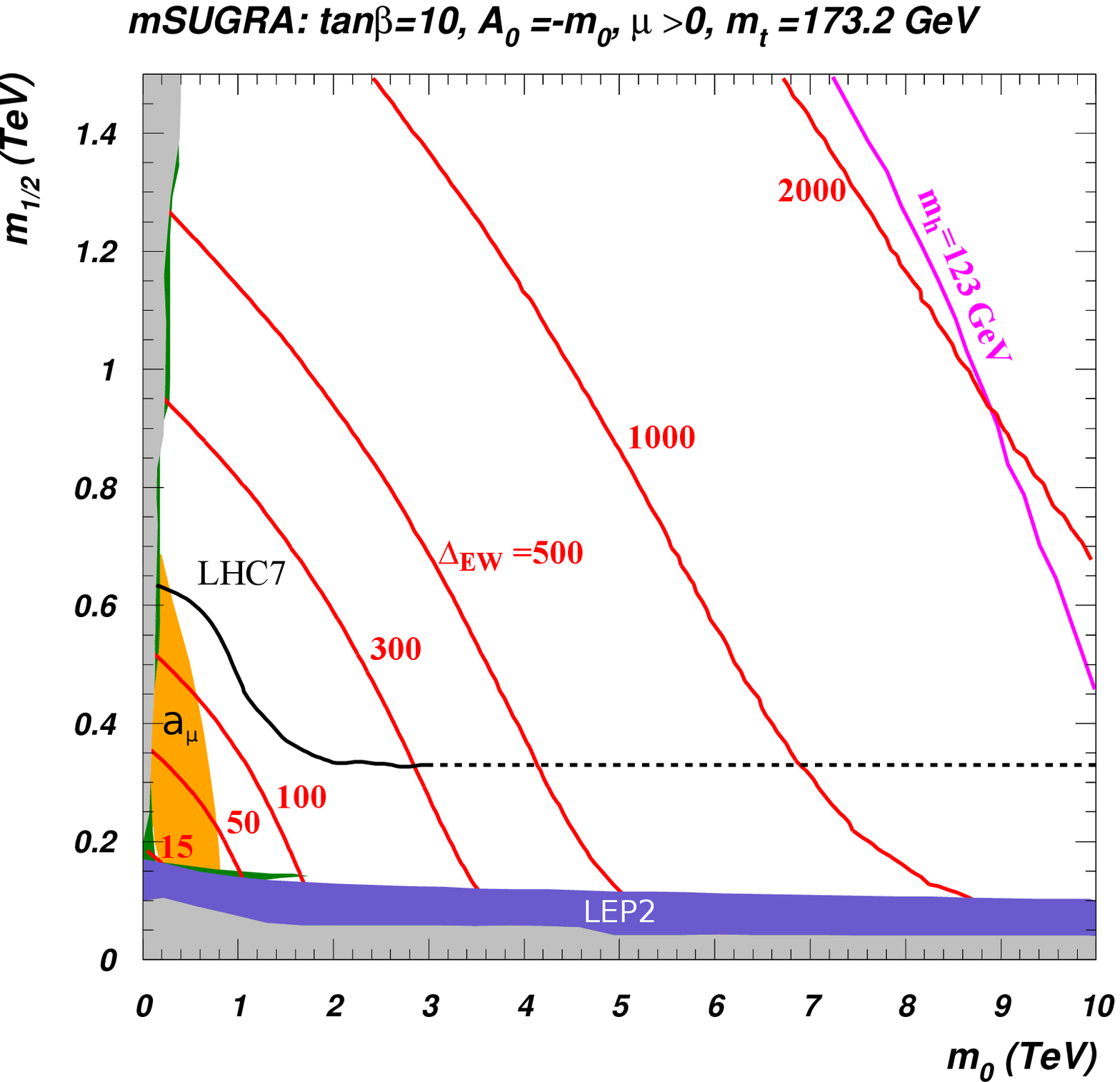}
\caption{Contours of {\it a}) $\Delta_{\rm HS}$ and {\it b}) $\Delta_{\rm EW}$ in the 
mSUGRA model with $A_0=-m_0$, $\tan\beta =10$ and $\mu>0$.
The value of $m_t$  as well as the various shaded/coloured regions are
as in Fig.~\ref{fig:tanb10_A0}.
}
\label{fig:tanb10_Am1}}

In Fig.~\ref{fig:tanb50_Am1}, we show the $m_0\ vs.\ m_{1/2}$ plane for
$A_0=-m_0$ but with $\tan\beta =50$. 
We see this is qualitatively very similar to the previous figure aside from the
sizeable LHCb excluded region on the low $m_0$ portion of the plane. 
Again the theoretically excluded region occurs at values of $m_0$ far 
beyond the range shown.
Note though that the contour of $m_h=123$~GeV has moved to slighly lower
$m_0$ values. Still, requiring $m_h>123$~GeV requires
$\Delta_{\rm HS}>5\times 10^3$, and $\Delta_{\rm EW}\agt 700$.
\FIGURE[tbh]{
\includegraphics[width=7cm]{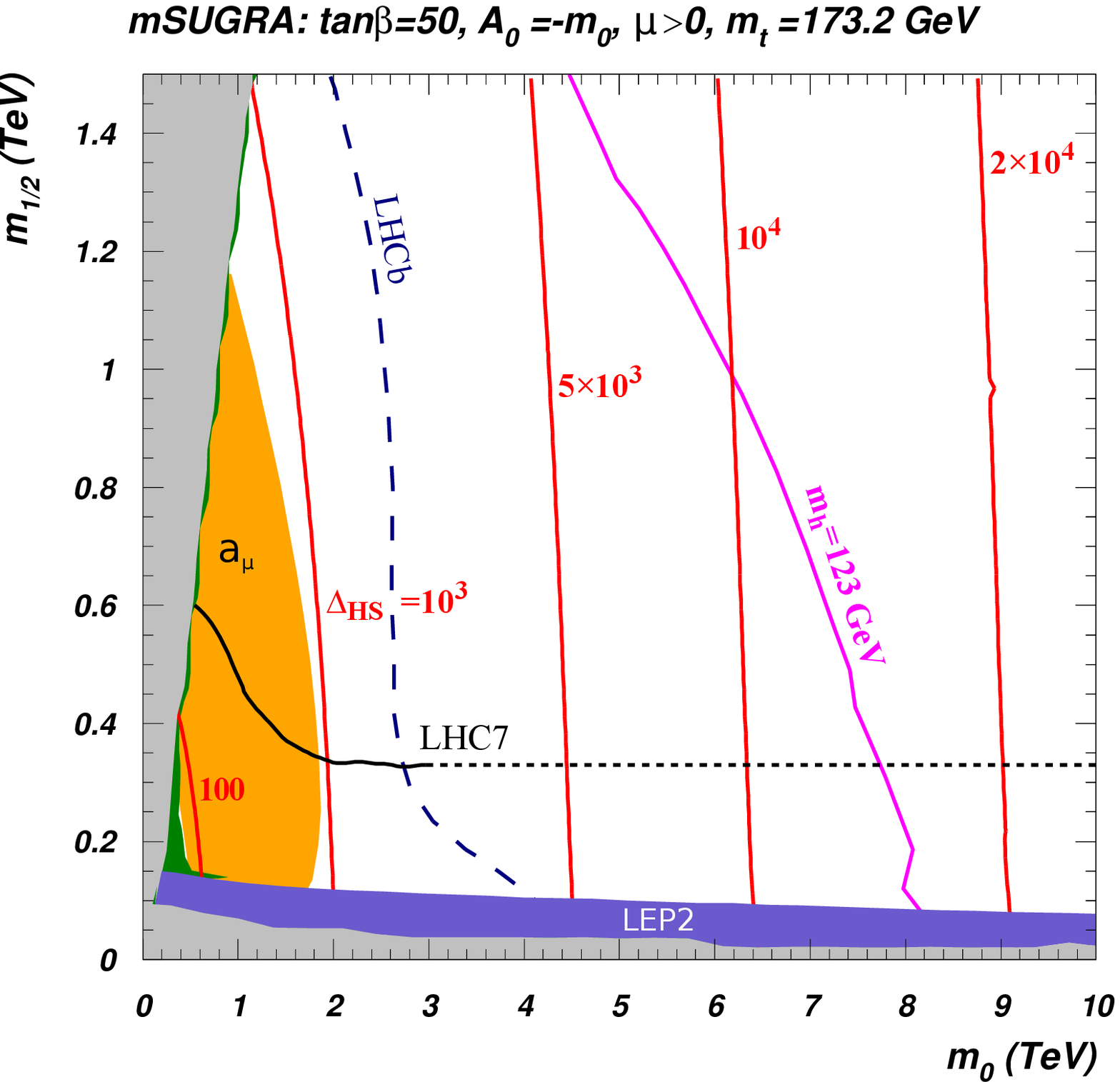}
\includegraphics[width=7cm]{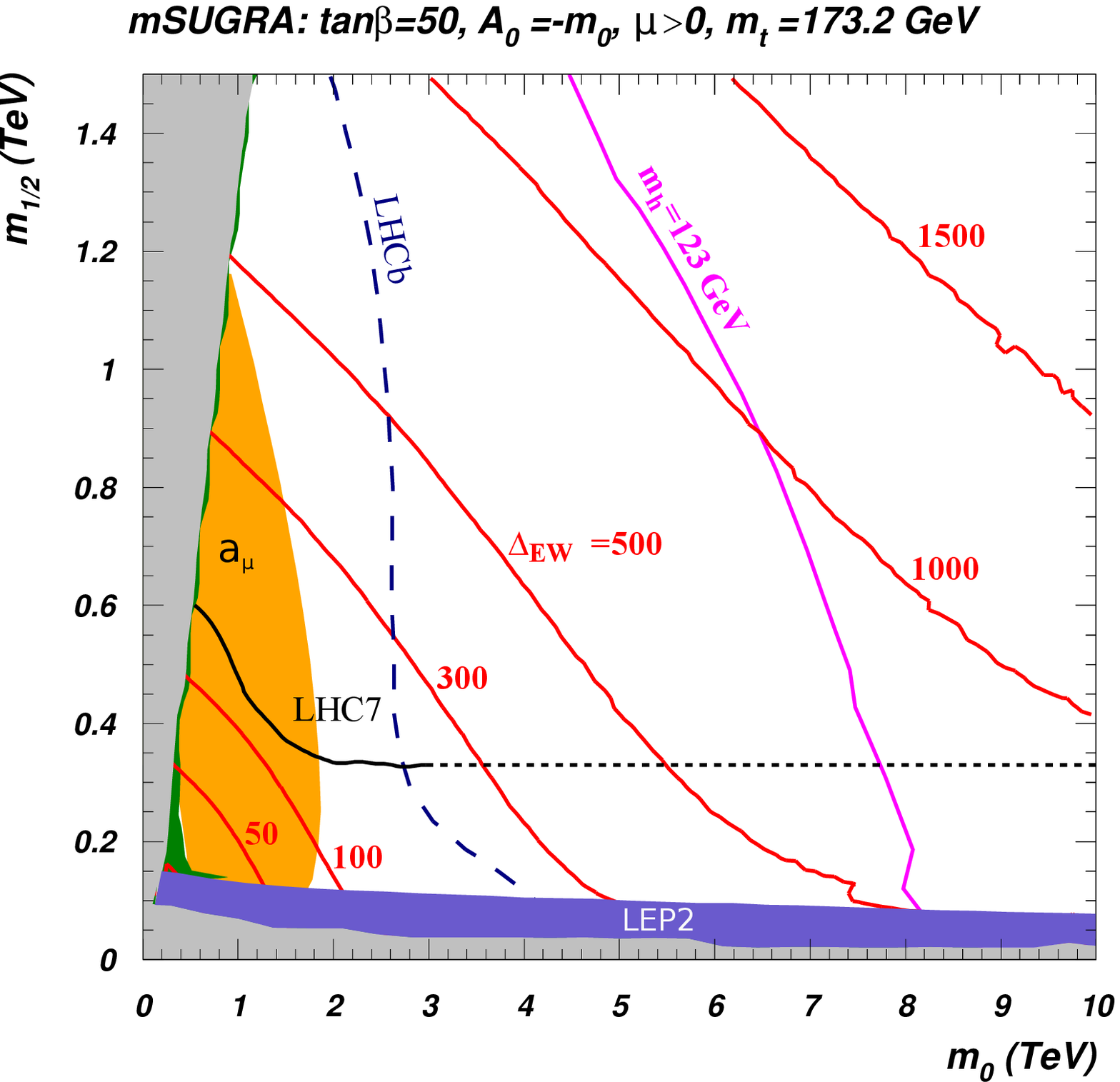}
\caption{Contours of {\it a}) $\Delta_{\rm HS}$ and {\it b}) $\Delta_{\rm EW}$
in the mSUGRA model with $A_0=-m_0$, $\tan\beta =50$ and $\mu>0$.  The
value of $m_t$ as well as the various shaded/coloured regions are as in
Fig.~\ref{fig:tanb50_A0}. }
\label{fig:tanb50_Am1}}

According to Ref.~\cite{bbm}, large mixing in the top squark sector and
consequently the largest values of $m_h$ occur in mSUGRA  for
$A_0\sim -2m_0$. In Fig.~\ref{fig:tanb10_Am2}, we show contours of
$\Delta_{\rm HS}$ and $\Delta_{\rm EW}$ for $\tan\beta =10$ and $A_0=-2m_0$.
We note again that the HB/FP region does not appear in this plane.
Notice  also that the
contours of $m_h=123$~GeV have moved all the way down to $m_0\sim 2$~TeV: 
thus, now much of the mSUGRA plane shown is {\it allowed} by the LHC
Higgs-like resonance discovery. In fact, the portion of the plane with
$m_0\agt 6-8$~TeV gives too large a value of $m_h>127$~GeV. The portion
of the $m_0\ vs.\ m_{1/2}$ plane allowed by both LHC7 sparticle searches
and by having $m_h\sim 123-127$~GeV requires $\Delta_{\rm HS}\sim
10^3-10^4$, or 0.1-0.01\% HSFT.  The EWFT required is $\Delta_{\rm EW}\agt
10^3$, also large. The lesson learned here is that
the remaining mSUGRA regions with $m_h\sim 123-127$~GeV, and which obey
sparticle mass constraints, are highly fine-tuned, even with  the less
restrictive EWFT measure.
\FIGURE[tbh]{
\includegraphics[width=7cm]{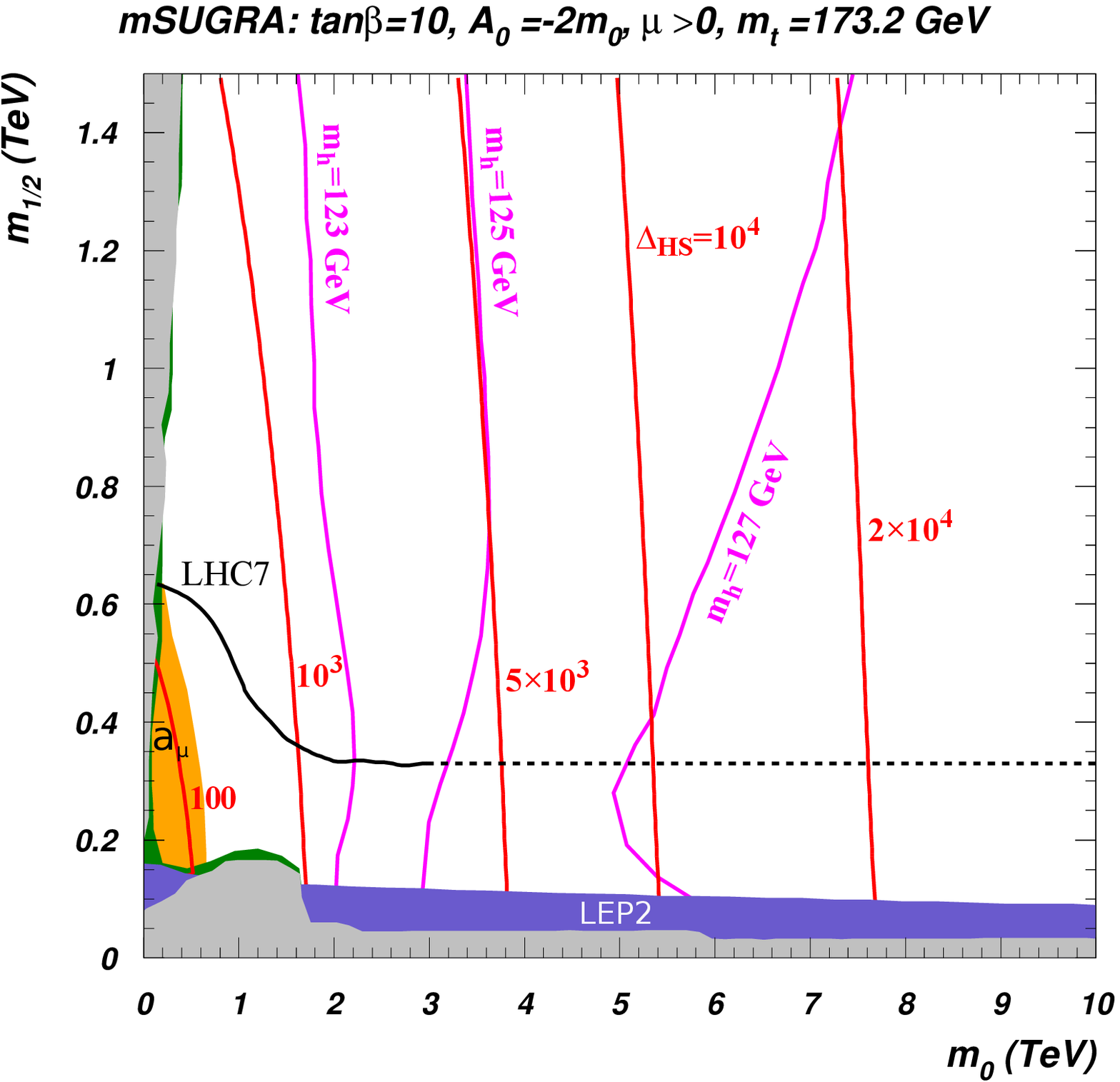}
\includegraphics[width=7cm]{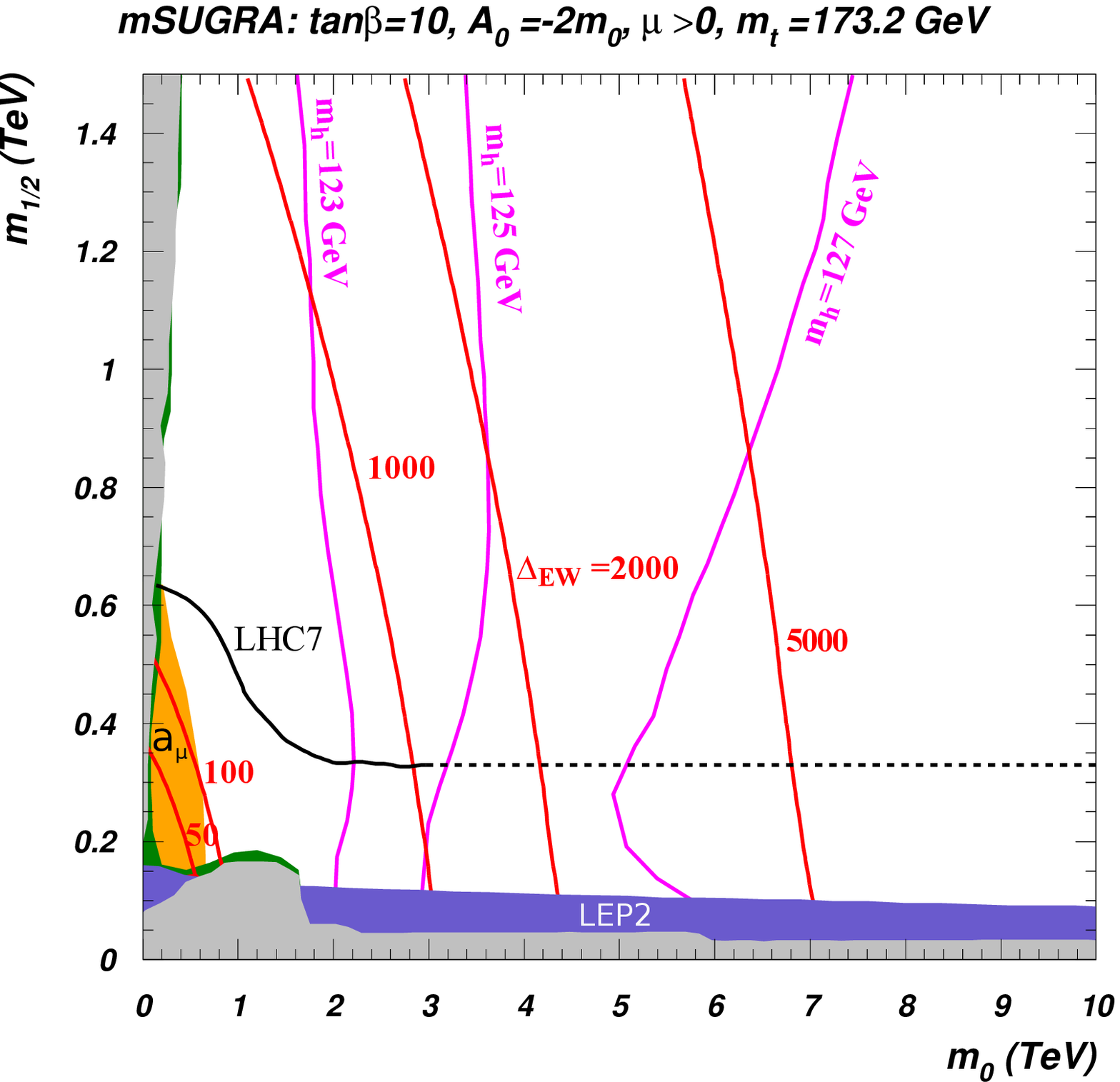}
\caption{Contours of {\it a}) $\Delta_{\rm HS}$ and {\it b}) $\Delta_{\rm EW}$
in the mSUGRA model with $A_0=-2m_0$, $\tan\beta =10$ and $\mu >0$. The
value of $m_t$ as well as the various shaded/coloured regions are as in
Fig.~\ref{fig:tanb10_A0}.}
\label{fig:tanb10_Am2}}

Fig.~\ref{fig:tanb50_Am2} shows the mSUGRA plane for $A_0=-2m_0$ but
with $\tan\beta =50$.  In this case, large theoretically excluded
parameter regions appear and these only grow larger until the
entire parameter space collapses for even higher $\tan\beta\sim
55-60$\cite{csaba}. The region on the right is forbidden because $m_A^2$ turns
negative, not because $|\mu|$ becomes small: this is why there is no
DM-allowed region for large values of $m_0$.  
The low $m_{1/2}$ and low $m_0$ portions of the plane marked LHCb 
are excluded due to too large a $B_s\to\mu^+\mu^-$ branching fraction.
The $m_h=123$~GeV contour nearly
coincides with $\Delta_{\rm HS}=10^3$ and $\Delta_{\rm EW}=500$. In this case,
values of $m_0\agt 6$~TeV are excluded as giving rise to too heavy a
value of $m_h$.  Thus, again the regions with $m_h\sim 123-127$~GeV and
obeying LHC7 sparticle search constraints, are highly fine-tuned.
\FIGURE[tbh]{
\includegraphics[width=7cm]{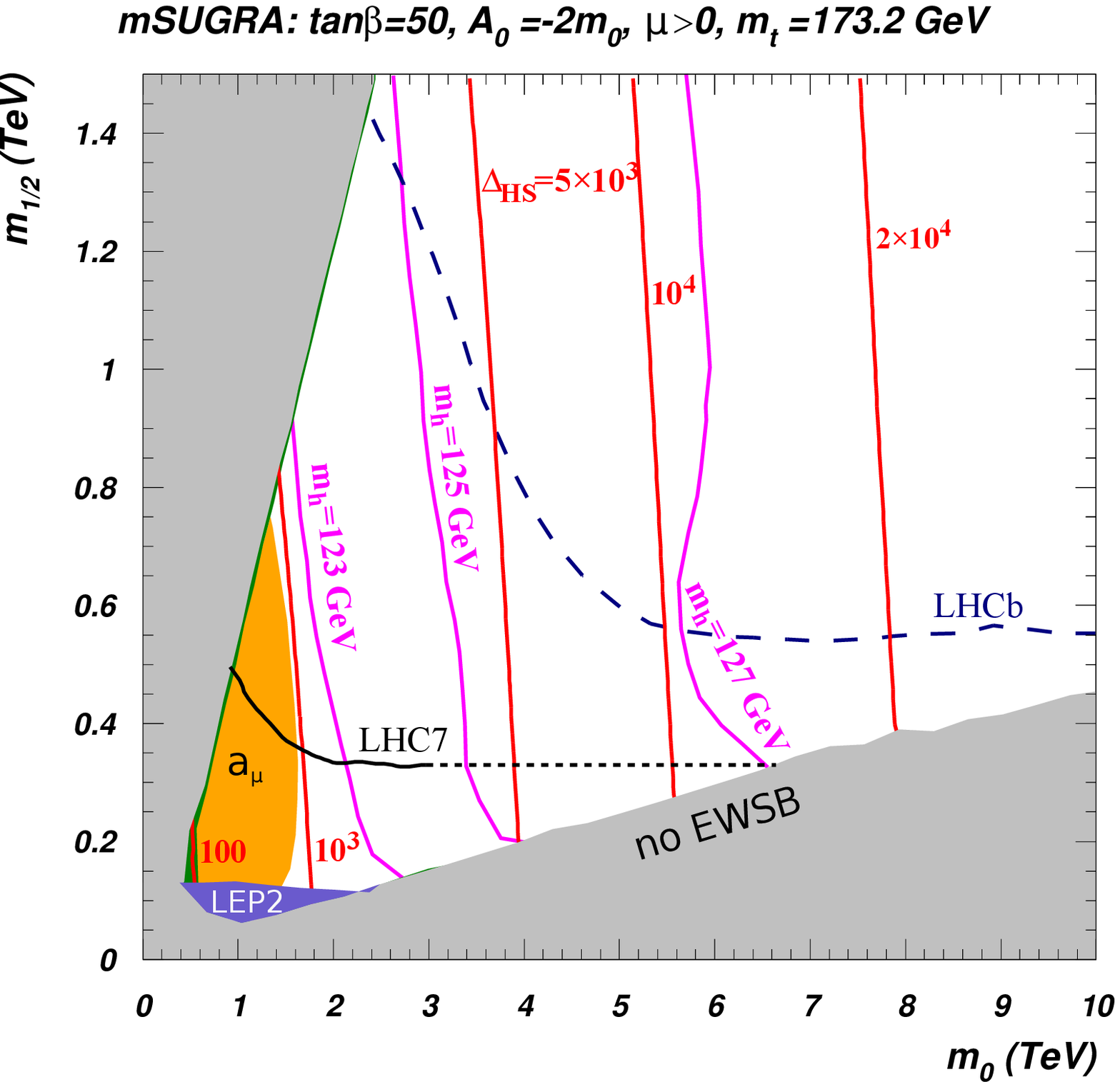}
\includegraphics[width=7cm]{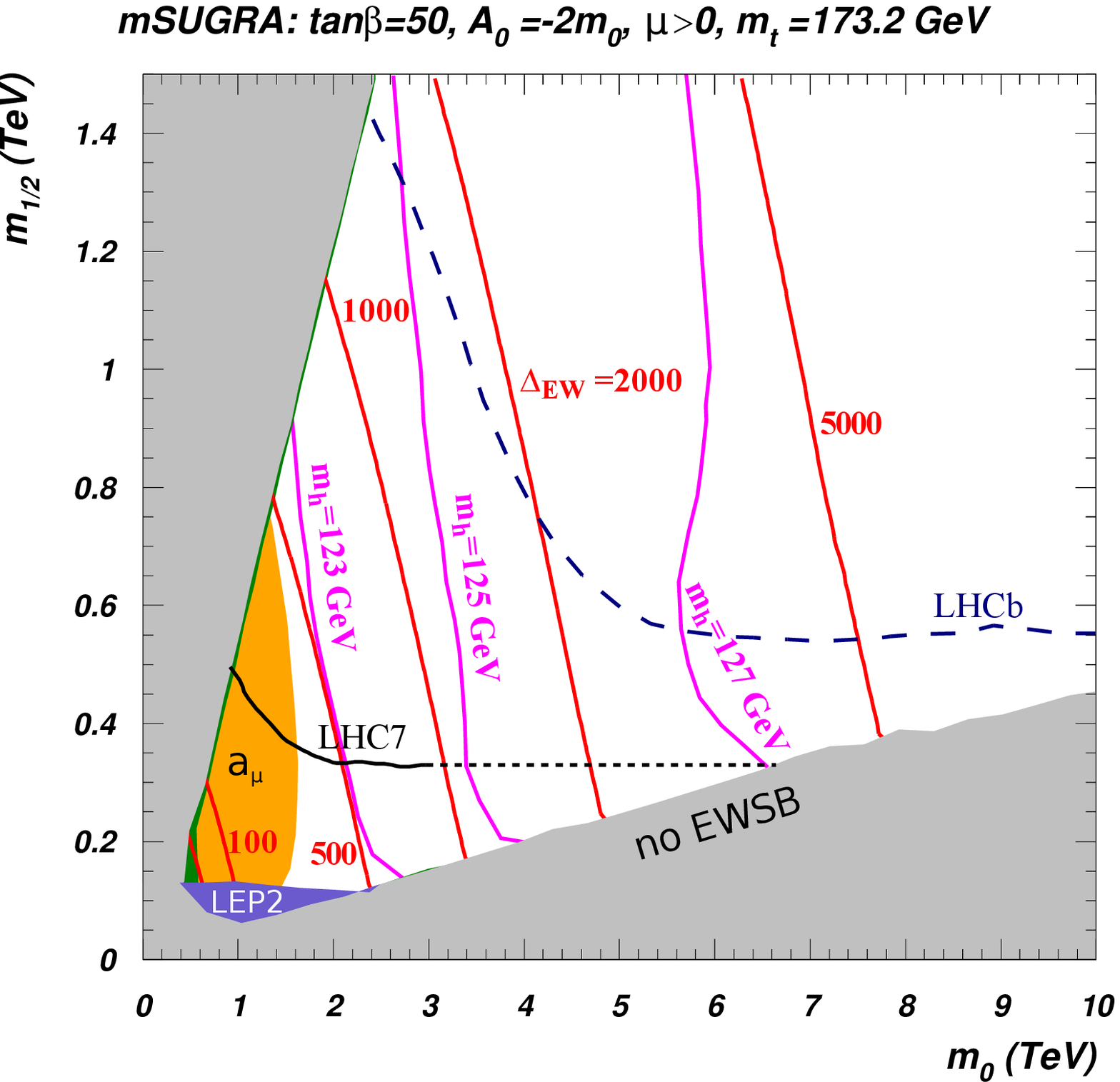}
\caption{Contours of {\it a}) $\Delta_{\rm HS}$ and {\it b}) $\Delta_{\rm EW}$
in the mSUGRA model with $A_0=-2m_0$, $\tan\beta =50$ and $\mu >0$. The
value of $m_t$ as well as the various shaded/coloured regions are as in
Fig.~\ref{fig:tanb50_A0}.}
\label{fig:tanb50_Am2}}

Before closing this section, we digress to compare our results for the 
EWFT measure with some results in the recent literature~\cite{many} for
the fine-tuning within the mSUGRA/CMSSM model calculated using the
procedure described at the end of Sec.~\ref{ssec:hsft}.  We have already
argued at the end of that section that the fine-tuning measure that
results from substituting $m_{H_u}^2$ in Eq.~(\ref{eq:otherft}) and the
analogous expression for $m_{H_d}^2$ into Eq.~(\ref{eq:mssmmu}) should
match our EWFT measure. To check this, we have compared our results in
Fig.~\ref{fig:tanb10_A0}{\it b}) to those in Fig.~1 of the first paper
of Ref.~\cite{many}. There, these authors show the {\it minimum value}
of their fine-tuning parameter $\Delta$ in the $m_0-m_{1/2}$ plane,
marginalizing over a range of $A_0$ and $\tan\beta$. We see that the
shapes of their $\Delta$ contours are qualitatively similar (except in
the large $m_0$ region where the contours turn around because radiative
correction effects are important) to those of
the contours in frame~{\it b}) of Figs.~\ref{fig:tanb10_A0} and
\ref{fig:tanb50_A0}.  We use our $A_0=0$ figures for this comparison
because of all the figures these have the smallest
value of $\Delta_{\rm EW}$.  We have also checked that for any chosen value
of $m_0$ and $m_{1/2}$ $\Delta$ of Antusch {\it et al.} has a magnitude
similar to (but never larger than) the corresponding lowest $\Delta_{\rm
EW}$ that we obtain for any choice of $A_0$ and $\tan\beta$.

%%%%%%%%%%%%%%%%%%%%%%%%%%%%%%%%%%%%%%%%%%%%%%%%%
\section{Scan over mSUGRA parameter space}
\label{sec:scan}
%%%%%%%%%%%%%%%%%%%%%%%%%%%%%%%%%%%%%%%%%%%%%%%%%

While the results of the previous section provide an
overview of both the EWFT and the HSFT measures
in  light of LHC7 and LHC8 constraints on sparticle and  Higgs boson masses,
we only presented results for particular choices of $A_0$ and $\tan\beta$, 
and for $\mu >0$.  In this Section, we present results from a scan
over the complete mSUGRA parameter space with the following range of model parameters:
\bea
m_0:\ 0-15\ {\rm TeV},\\
m_{1/2}:\  0-2\ {\rm TeV},\\
-2.5<\ A_0/m_0\ <2.5,\\
\tan\beta :\ 3-60 .
\label{eq:pspace}
\eea
We will show results for both $\mu >0$ and $\mu <0$.
For each solution generated, we require 
\begin{enumerate}
\item electroweak symmetry be radiatively broken (REWSB), 
\item the neutralino $\tz_1$ is the lightest MSSM particle, 
\item the light chargino mass obeys the LEP2 limit that
$m_{\tw_1}>103.5$~GeV~\cite{lep2},
\item $m_h=125\pm 2$~GeV in accord with the recent Higgs-like
resonance discovery at LHC~\cite{atlas_h,cms_h},
\item the calculated value of $BF(B_s\to \mu^+\mu^- )$ lie within
$(2-4.7)\times 10^{-9}$ in accord with recent LHCb measurements~\cite{lhcb} and 
\item the mass spectra obey LHC7 sparticle mass constraints for the
mSUGRA model~\cite{atlas_susy,cms_susy}.
\end{enumerate}

Our first results are shown in Fig.~\ref{fig:scan_m0} for {\it
a}) $\Delta_{\rm HS}$ and {\it b}) $\Delta_{\rm EW}$ versus
$m_0$. Solutions with $\mu <0$ are shown as blue circles while solutions
with $\mu >0$ are shown in red crosses. 
Note that here, and in subsequent figures, there are many points for $\mu <0$ (red circles) that are not
visible as these are covered by the red crosses for $\mu >0$.
In frame {\it a}), we see that
$\Delta_{\rm HS}$ values occupy a rather narrow band which increases
monotonically with $m_0$.  Values of $m_0\alt 1$~TeV are excluded by the
requirement $m_h > 123$~GeV. The $\mu >0$ and $\mu <0$ solutions occupy
essentially the same band. This is not surprising because the large
logarithms are essentially independent of the sign of $\mu$. The minimum
allowed value of $\Delta_{\rm HS}$ is $\sim 1000$, so that at least
0.1\% fine-tuning is required of all remaining mSUGRA solutions. The
minimum for $\Delta_{\rm HS}$ occurs at $m_0\sim 1500$~GeV. This minimal
$\Delta_{\rm HS}$ solution is shown as a benchmark point in
Sec.~\ref{sec:bm}. For $m_0$ as high as 15~TeV, $\Delta_{\rm HS}$ increases
to nearly $10^5$.  In frame {\it b}), we show $\Delta_{\rm EW}$ versus
$m_0$. Here, the shape of the allowed region is very different from the
$\Delta_{\rm HS}$ case in frame {\it a}). Low values of $m_0$ can give
$m_h > 123$~GeV only if $|A_0/m_0|$ is sizeable and, as we have already
seen, yield $\Delta_{\rm EW}$ of at least several hundred. Smaller
values of $\Delta_{\rm EW}$ are obtained only in the HB/FP region where
$m_0$ is large. In other words, in the ``hole region'' in frame {\it b}), 
we have $m_h < 123$~GeV.  The point with the minimum value of
$\Delta_{\rm EW}\sim 100$ occurs at $m_0\sim 7900$~GeV, and is shown as the
electroweak benchmark point in the next section. Over the remaining
mSUGRA parameter space, at best 1\% EWFT is required.
\FIGURE[tbh]{
\includegraphics[width=7cm]{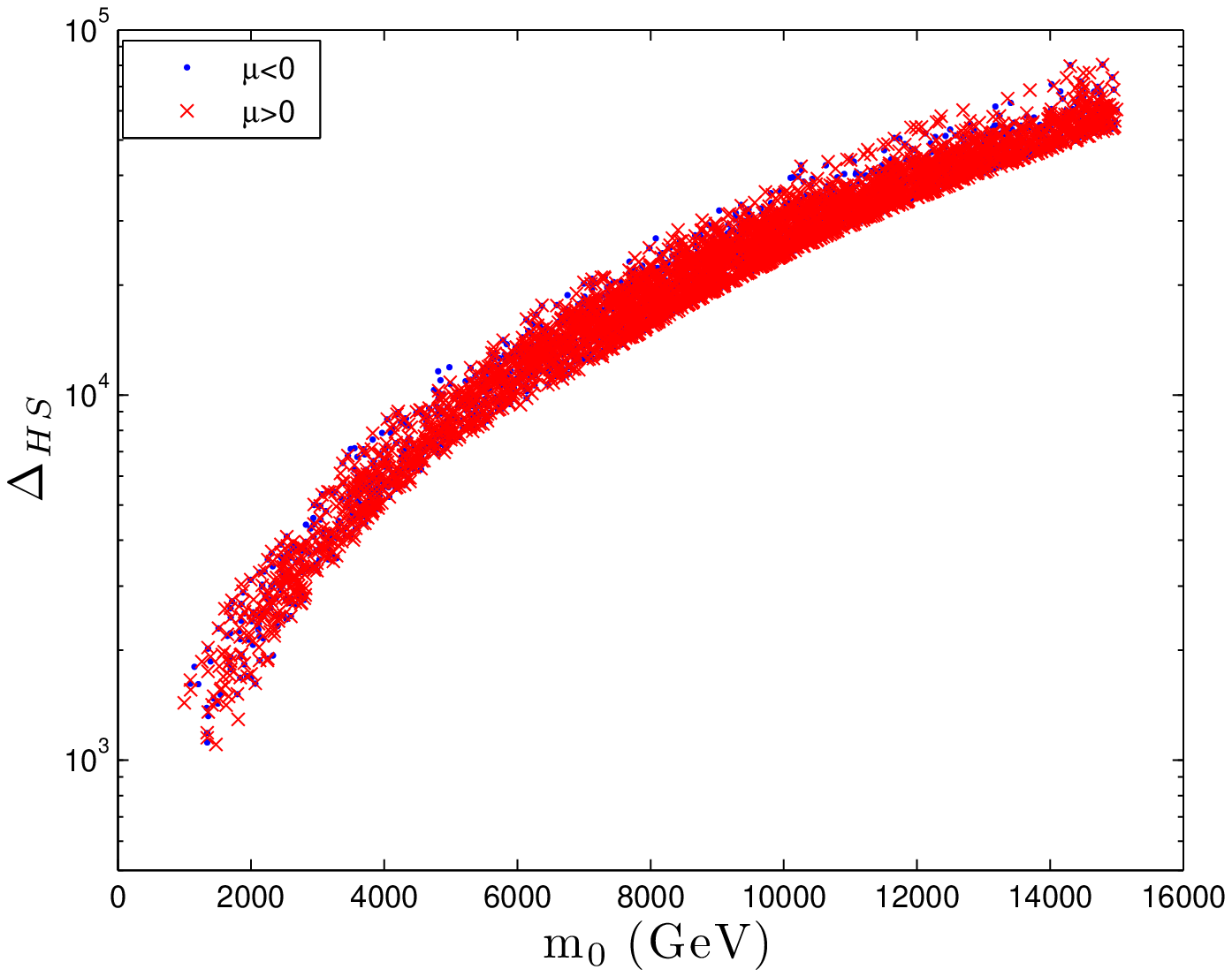}
\includegraphics[width=7cm]{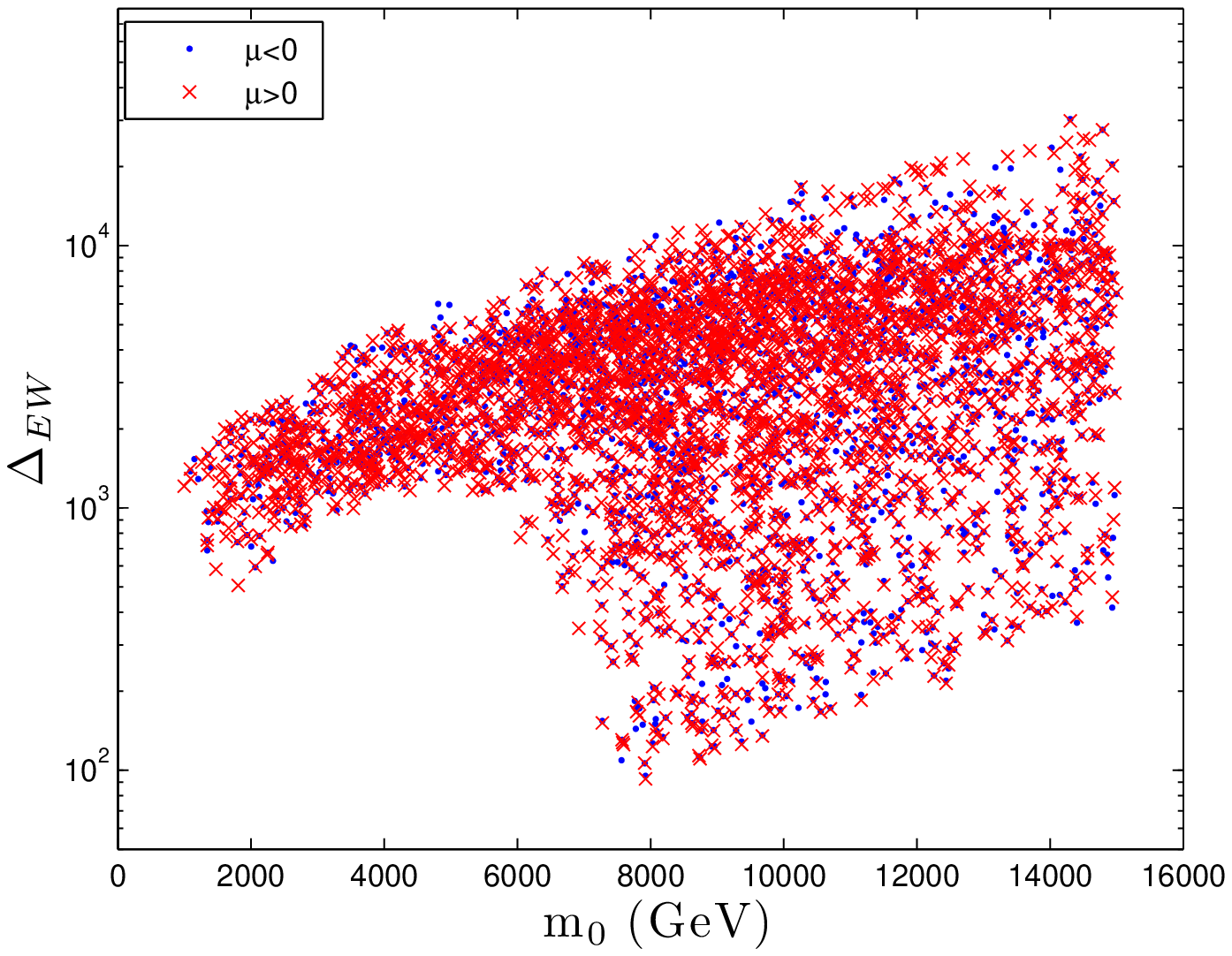}
\caption{Fine-tuning measures $\Delta_{\rm HS}$ and $\Delta_{\rm EW}$
versus $m_0$ from a scan over mSUGRA/CMSSM parameter space for $\mu >0$
(red crosses) and $\mu<0$ (blue circles).  We take $m_t=173.2$~GeV.  }
\label{fig:scan_m0}}

In Fig.~\ref{fig:scan_mhf}, we show the distributions of $\Delta_{\rm
HS}$ and $\Delta_{\rm EW}$ versus $m_{1/2}$.  The sharp edge on the left
is a reflection of the lower limit on $m_{\tg}$ from LHC7 searches. 
In frame {\it a}) for $\Delta_{\rm HS}$, 
we see that the minimal $\Delta_{\rm HS}$ is spread
across a wide spectrum of $m_{1/2}$ values. This is consistent with the
behavior of $\Delta_{\rm HS}$ shown in Fig.~\ref{fig:tanb10_A0}{\it
a}) where the HSFT contours are nearly vertical, indicating little
dependence on $m_{1/2}$. In frame {\it b}), the minimal values of
$\Delta_{\rm EW}$ are also spread across the $m_{1/2}$ range.
For both $\Delta_{HS}$ and $\Delta_{EW}$, there may be a  slight 
preference for lower $m_{1/2}$ values.
\FIGURE[tbh]{
\includegraphics[width=7cm]{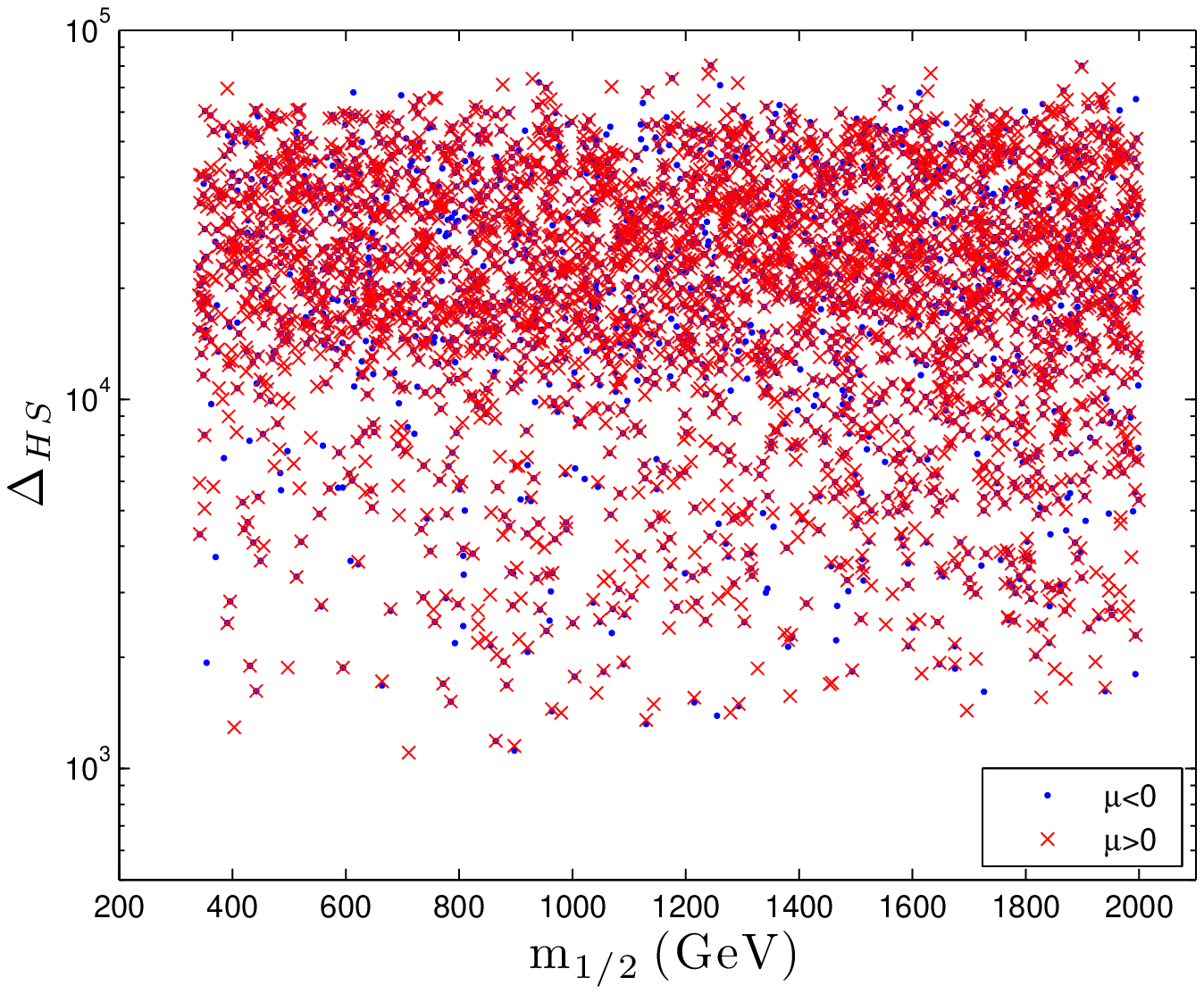}
\includegraphics[width=7cm]{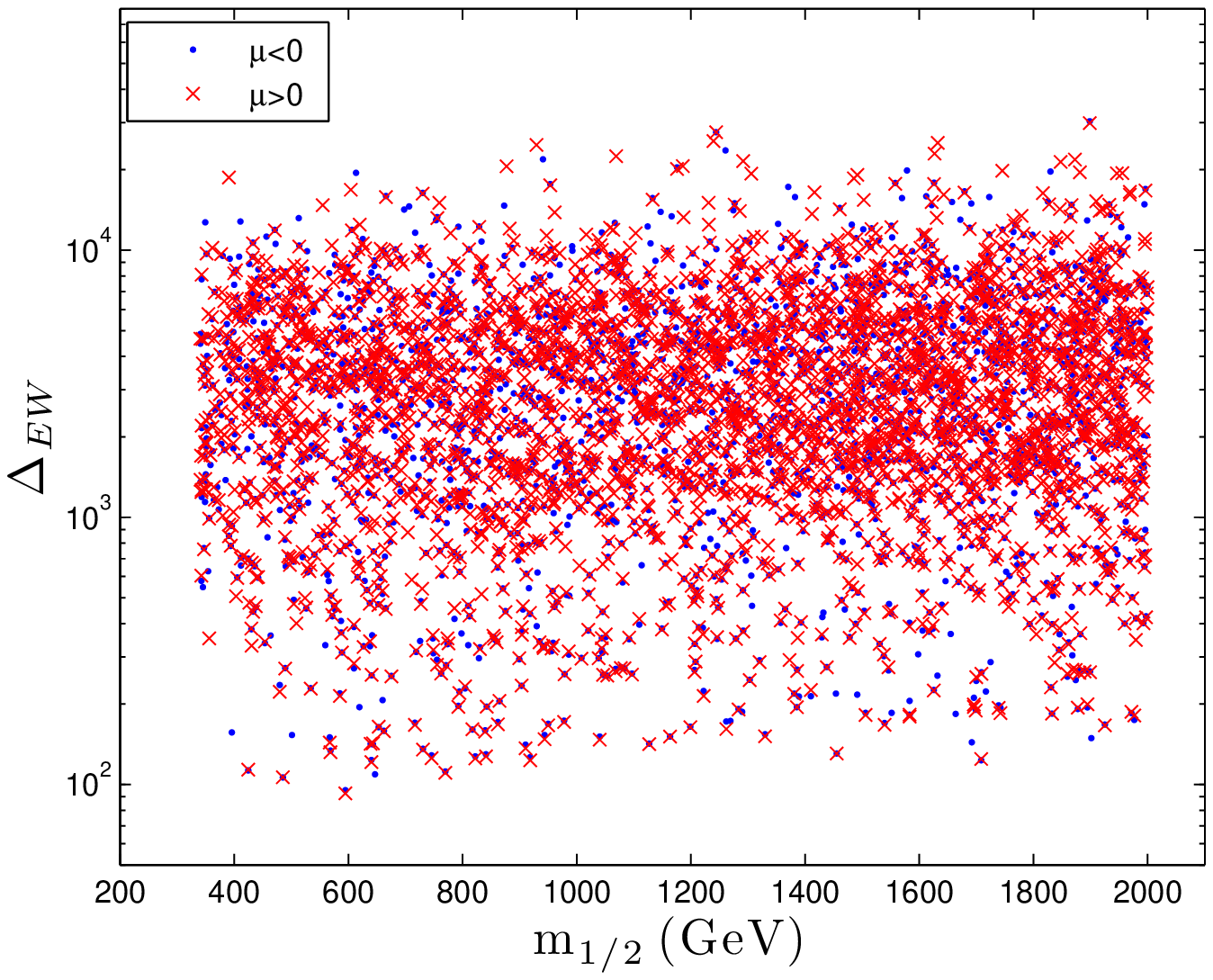}
\caption{Fine-tuning measures $\Delta_{\rm HS}$ and $\Delta_{\rm EW}$ 
versus $m_{1/2}$ from a scan over
mSUGRA/CMSSM parameter space for $\mu >0$ (red crosses) and $\mu<0$
(blue circles).
}
\label{fig:scan_mhf}}

In Fig.~\ref{fig:scan_a0} we show how $\Delta_{\rm HS}$ and $\Delta_{\rm
  EW}$ are distributed versus $A_0/m_0$.  In frame {\it a}), we see
  that minimal $\Delta_{\rm HS}\sim 1000$ is obtained for $A_0/m_0\sim -2$,
  which is also the vicinity of where $m_h$ is maximal for given $m_0$
  and $m_{1/2}$ values.  There is also a minimum at $A_0/m_0\sim 2.5$,
  with $\Delta_{\rm HS}$ reaching only to $\sim 3000$~\footnote{The
  asymmetry of the minimum of $\Delta_{\rm HS}$ with respect to the sign
  of $A_0$ may only be a reflection of the fact that it is more
  difficult to generate large values of $m_h$ for positive values of
  $A_0$.}.  In frame {\it b}), the value of $\Delta_{\rm EW}$ is even more
  correlated with $A_0/m_0$.  For $|A_0/m_0|\alt 1$, $\Delta_{\rm EW}$
  tends to be smaller than for larger values of $|A_0/m_0|$. The
  solutions with the least EWFT occur at $A_0/m_0\sim \pm 0.6$, with the
  minimal $\Delta_{\rm EW}\sim 100$.  Once again, this occurs in the HB/FP
  region mentioned above. For larger magnitudes of $A_0/m_0$, the HB/FP region is absent, 
  and $\Delta _{\rm EW}$ is much larger. 
  The  gap in the plots around $A_0/m_0\sim 0$ occur because it is nearly
  impossible to generate $m_h$ as heavy as $123-127$~GeV for such low
  values of trilinear couplings\cite{bbm}.
\FIGURE[tbh]{
\includegraphics[width=7cm]{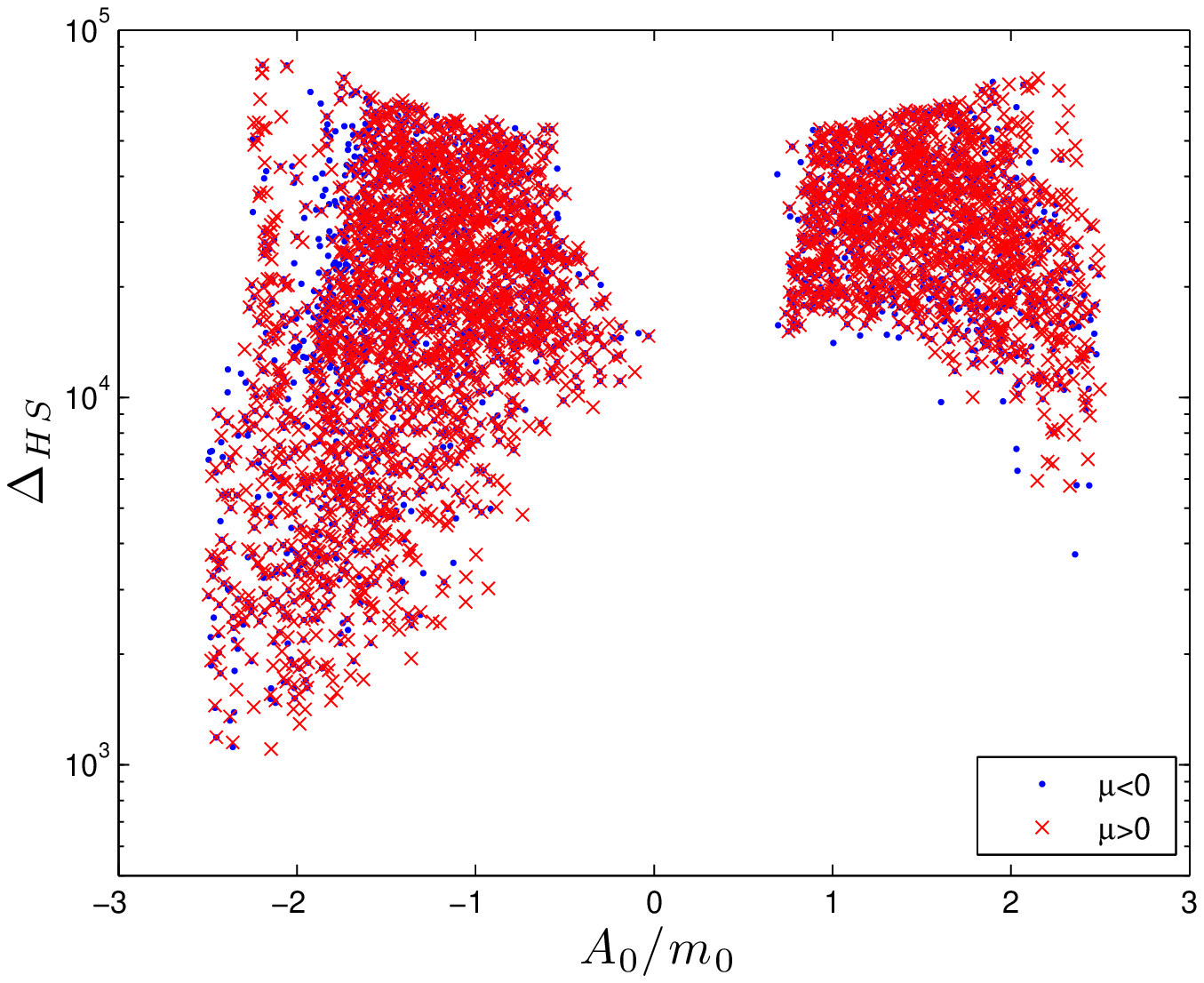}
\includegraphics[width=7cm]{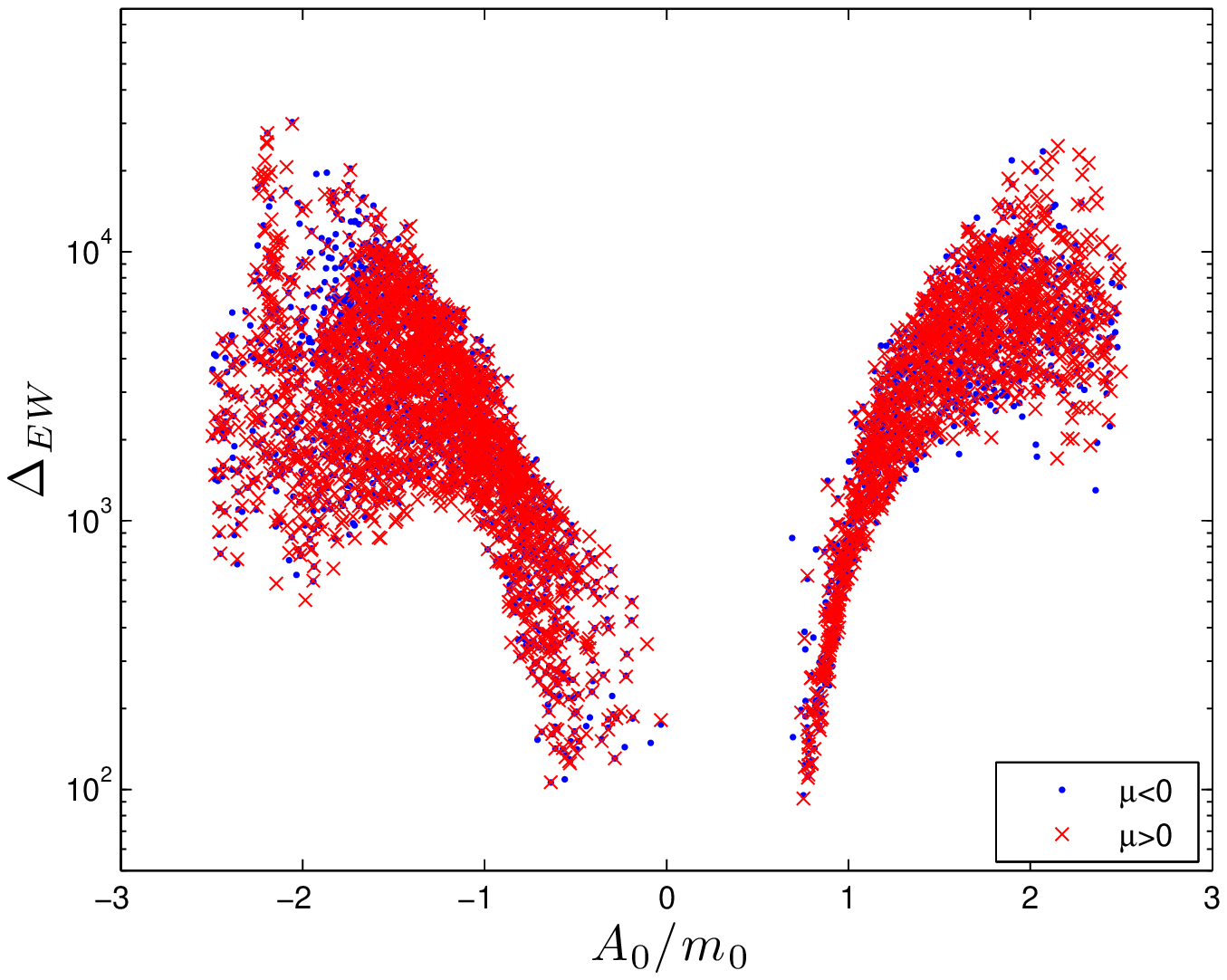}
\caption{Fine-tuning  measures $\Delta_{\rm HS}$ and $\Delta_{\rm EW}$ 
versus $A_0/m_0$ from a scan over
mSUGRA/CMSSM parameter space for $\mu >0$ (red crosses) and $\mu<0$
(blue circles).
}
\label{fig:scan_a0}}

In Fig.~\ref{fig:scan_tanb}, we plot $\Delta_{\rm HS}$ and $\Delta_{\rm EW}$
versus $\tan\beta$.  The minimal $\Delta_{\rm HS}$ and $\Delta_{\rm EW}$
solutions are spread uniformly across a range of $\tan\beta$ values. At
very low $\tan\beta\alt 10$ values, it is difficult to generate solutions with 
$m_h\agt 123$~GeV unless mSUGRA parameters are extremely large, leading to
high fine-tuning.
\FIGURE[tbh]{
\includegraphics[width=7cm]{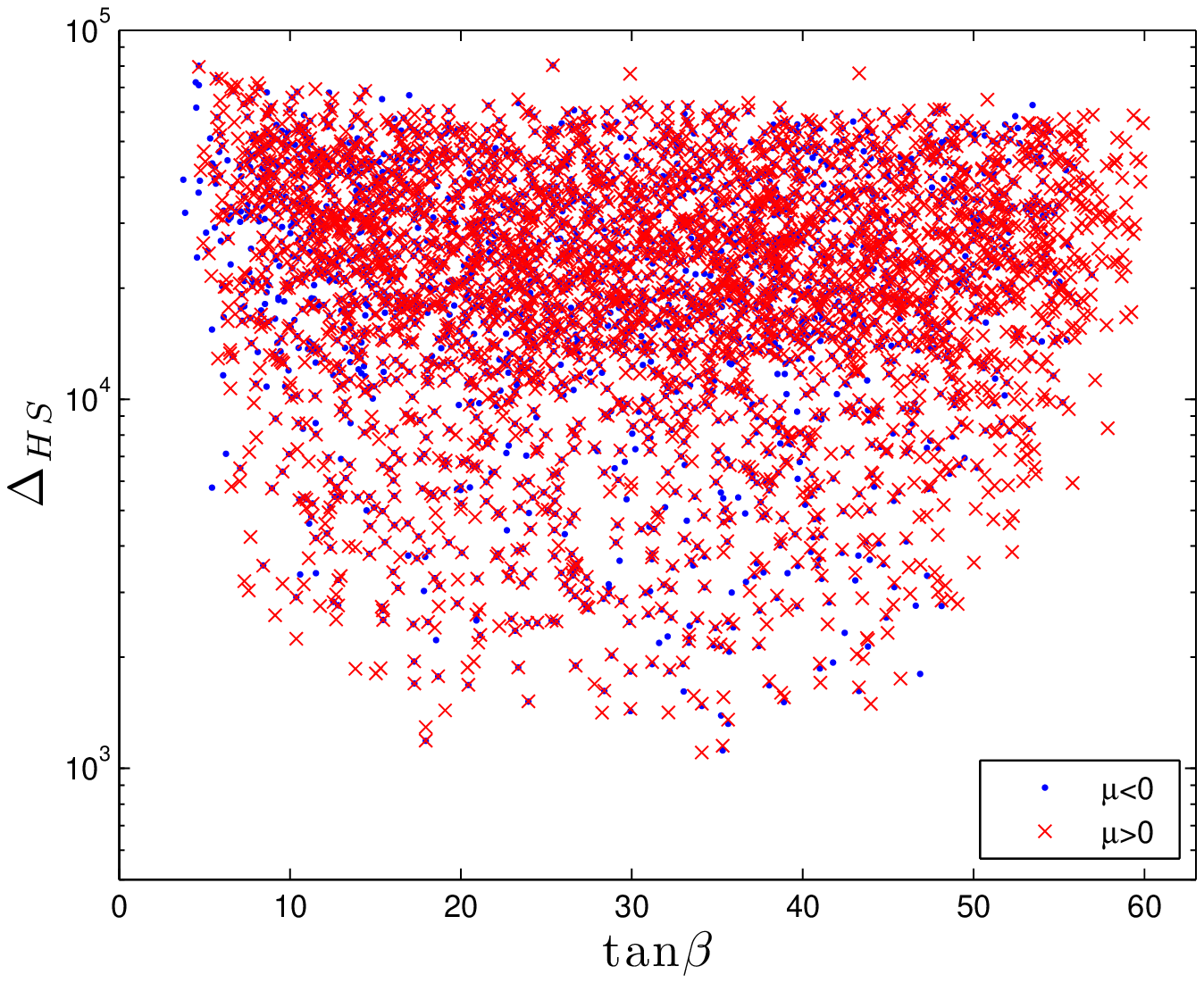}
\includegraphics[width=7cm]{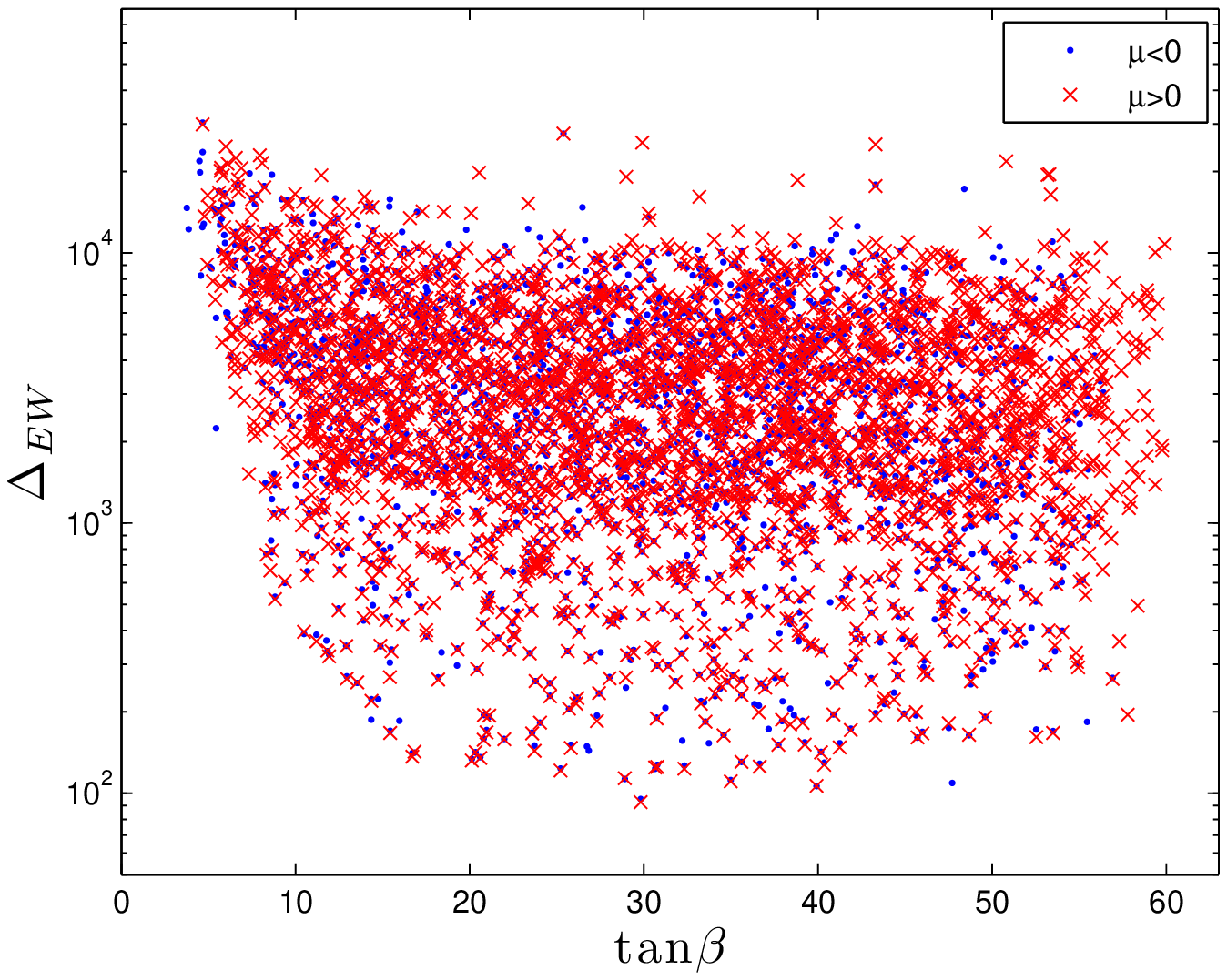}
\caption{Fine-tuning measures $\Delta_{\rm HS}$ and $\Delta_{\rm EW}$ 
versus $\tan\beta$ from a scan over
mSUGRA/CMSSM parameter space for $\mu >0$ (blue circles) and $\mu<0$
(red crosses).
}
\label{fig:scan_tanb}}
%

%%%%%%%%%%%%%%%%%%%%%%%%%%%%%%%%%%%%%%%%%%%%%%%%%
\section{Lowest fine-tuning mSUGRA benchmarks}
\label{sec:bm}
%%%%%%%%%%%%%%%%%%%%%%%%%%%%%%%%%%%%%%%%%%%%%%%%%

What is apparent from our results so far is that, after imposing LHC7
sparticle mass constraints and requiring that $m_h=125\pm 2$~GeV on the
mSUGRA/CMSSM model, the viable solutions are fine-tuned to at least 1\%
even with the less stringent EWFT measure. With a fine-tuning
measure that knows about the high scale origin of mSUGRA parameters, the 
required fine-tuning is increased by an order of magnitude. 
Nonetheless, our understanding of how SUSY breaking parameters arise is 
extremely limited and it
remains possible that nature may appear fine-tuned to a certain degree.
With this in mind, we exhibit and qualitatively examine  the
features of the lowest  $\Delta_{HS}$ and the lowest $\Delta_{\rm  EW}$ 
solutions in the mSUGRA/CMSSM framework. These are listed in
Table~\ref{tab:bm} as solutions HS1 and EW1.
\begin{table}\centering
\begin{tabular}{lcc}
\hline
parameter & HS1 & EW1 \\
\hline
$m_0$      & 1472.0 & 7926.4  \\
$m_{1/2}$   & 711.0 & 594.6  \\
$A_0$      & -3157.4 & 5968.2  \\
$\tan\beta$& 34.1 & 29.8  \\
\hline
$m_{\tg}$   & 1662.5 & 1589.9   \\
$m_{\tu_L}$ & 2058.8 & 7949.5  \\
$m_{\tu_R}$ & 2025.4 & 7972.3  \\
$m_{\te_R}$ & 1494.7 & 7922.0  \\
$m_{\tst_1}$& 887.8 & 4547.6  \\
$m_{\tst_2}$& 1499.8 & 6197.4  \\
$m_{\tb_1}$ & 1475.6 & 6175.2  \\
$m_{\tb_2}$ & 1731.0 & 7406.6  \\
$m_{\ttau_1}$ & 1023.9 & 7187.3  \\
$m_{\ttau_2}$ & 1347.7 & 7563.8 \\
$m_{\tnu_{\tau}}$ & 1339.9 & 7565.6  \\
$m_{\tw_2}$ & 1550.1  & 657.6  \\
$m_{\tw_1}$ & 594.0  & 490.4  \\
$m_{\tz_4}$ & 1547.9 & 659.0 \\ 
$m_{\tz_3}$ & 1545.2 & 638.5 \\ 
$m_{\tz_2}$ & 591.9 & 487.7 \\ 
$m_{\tz_1}$ & 308.1 & 257.6 \\ 
$m_h$       & 123.2 & 123.1  \\ 
$\mu$      & 1550.8 & 619.7  \\
$m_A$       & 1626.8 & 6682.5  \\ 
\hline
$\Omega_{\tz_1}^{th}h^2$ & 12.3 & 9.4 \\
$BF(b\to s\gamma)\times 10^4$ & $2.7$  & 3.1 \\
$BF(B_s\to \mu^+\mu^-)\times 10^9$ & $4.4$  & 3.8  \\
$\sigma^{SI}(\tz_1 p)$ (pb) & $1.4\times 10^{-11}$  & $1.6\times 10^{-10}$ \\
$\Delta_{\rm HS}$ & 1105 & $1.5\times 10^4$ \\
$\Delta_{\rm EW}$ & 582.9 & 92.4 \\
\hline
\end{tabular}
\caption{Input parameters and masses in~GeV units
for the two mSUGRA/CMSSM benchmark points with the lowest values of
$\Delta_{\rm HS}$ and $\Delta_{\rm EW}$   after imposing
$m_h=125\pm 2$~GeV and also the LHC7 sparticle mass bounds. 
We take $m_t=173.2$~GeV.
}
\label{tab:bm}
\end{table}

Solution HS1 has $\Delta_{\rm HS}=1100$ and so requires $\sim 0.1\%$
fine-tuning.  The EWFT parameter $\Delta_{\rm EW}\sim 600$, requiring
$\sim 0.2\%$ fine-tuning.  HS1 has $m_0\sim 1500$~GeV, lying at the lower
edge of the band of solutions shown in Fig.~\ref{fig:scan_m0}. With
$m_{\tg}\sim 1660$~GeV, and $m_{\tq}\sim 2000$~GeV, this solution lies
beyond the reach of LHC8 searches with up to 30~fb$^{-1}$\cite{update},
but should be accessible to LHC14 searches with $\sim 10-20$~fb$^{-1}$~\cite{lhcreach}.  
The relatively light top squarks allow for
$\tg\to t\tst_1$ decay at $\sim 100\%$, followed by $\tst_1\to
t\tz_1$. Thus, gluino pair production will give rise to
$t\bar{t}t\bar{t}+\eslt$ events at LHC and may be searchable even in
the multi-jet plus $\eslt$ channel\cite{jason}. First generation squark
pair production and corresponding $\tq\tg$ production will augment this
rate since typically $\tq\to q\tg$ for first and second
generation squarks. Production of second and third generation squarks
will be suppressed by parton distribution functions.  The HS1 solution
has $\Omega_{\tz_1}^{th}h^2\sim 12$, so would produce too many
neutralinos in the early universe under the standard cosmology. Late
time entropy production\cite{bls} or neutralino decay to a lighter state, {\it
e.g.}  $\gamma +axino$ in extended models\cite{bbs}, can bring such a model into
accord with the measured relic abundance. The $b\to s\gamma$ branching
fraction is somewhat below measured values, although additional
flavor-violating Lagrangian soft terms could bring this value into
accord with measurements without affecting LHC phenomenology.

The solution EW1 has $\Delta_{\rm HS}\sim 1.5\times 10^4$, but
$\Delta_{\rm EW}\sim 100$ so that the latter requires EWFT at the 1\%
level. The reader may wonder whether it makes sense to talk about low
values of $\Delta_{\rm EW}$ when $\Delta_{\rm HS}$ is so much larger. In
this connection, it may be worth allowing for the possibility that the
mSUGRA framework may itself one day be derived from an underlying theory
along with specific relations between seemingly unrelated mSUGRA
parameters that lead to cancellations of the terms containing the large
logarithms, as discussed at the end of Sec.~\ref{ssec:hsft}.  Returning
to the EW1 point in the Table with $m_{\tg}\sim 1600$~GeV and
$m_{\tq}\sim 6-8$~TeV, we see that this model is only accessible to
LHC14 searches with $\sim 50-100$~fb$^{-1}$ of integrated luminosity\cite{lhcreach}. 
In this case, gluino pair production would be followed by gluino three-body
decays to multi-jet plus multi-lepton plus $\eslt$ final states. The
final states would be rich in $W$ and $Z$ bosons, leading to distinctive
signatures~\cite{btw_z}. The thermally-produced neutralino abundance
$\Omega_{\tz_1}^{th}h^2\sim 10$, so again a non-standard cosmology as
well as an extension of the spectrum is needed to bring this solution in
accord with the measured dark matter density. 

Both HS1 and EW1 points will need yet other new physics to bring them in
accord with the E821 measurement~\cite{bnl} of the muon magnetic moment
if this discrepancy continues to hold up.

%%%%%%%%%%%%%%%%%%%%%%%%%%%%%%%%%%%%%%%%%%%%%%%%%
\section{Concluding Remarks}
\label{sec:conclude}
%%%%%%%%%%%%%%%%%%%%%%%%%%%%%%%%%%%%%%%%%%%%%%%%%

The recent discovery of a 125~GeV Higgs-like resonance at LHC has set a
strong new constraint on supersymmetric models. In addition, the lack of
evidence for a SUSY signal at LHC now requires masses of strongly
interacting sparticles in models such as mSUGRA/CMSSM to be above the
1~TeV scale.  If LHC searches for sparticles continue without a new
physics signal, then the little hierarchy problem -- how to reconcile
the $Z$ and Higgs boson mass scale with the scale of SUSY breaking --
will become increasingly acute in models such as mSUGRA.\footnote{We do note that
the little hierarchy problem may be solved within the context of the MSSM 
if we go to non-universal SUGRA models: see {\it e.g.} Ref.~\cite{bbhmt,rns}.
Alternatively, invoking extra singlets\cite{nmssm} or extra vector-like matter\cite{vlikematter}
may provide additional contributions to $m_h$ while maintaining light top squarks
which seem to be required for low electroweak fine-tuning.}

In this paper, we have reported on results from the calculation of two
measures of fine-tuning in the mSUGRA/CMSSM model. The first --
$\Delta_{\rm HS}$ which includes information about the high scale origin
of mSUGRA parameters -- is the more stringent one. The second,
$\Delta_{\rm EW}$, depends only on the physical spectrum and couplings,
and so is universal to all models that yield the same weak scale
Lagrangian.  Our results incorporate the latest constraints from LHC7
sparticle searches along with a light Higgs scalar with $m_h\sim
123-127$~GeV.  We find $\Delta_{\rm HS}\agt 10^3$, or at best $0.1\%$
fine-tuning.  The more model-independent EWFT gives a $\Delta_{\rm
EW}\agt 10^2$, or at best 1\% fine-tuning.  The minimum value of
$\Delta_{EW}$ tends to occur near the FP region which extends to
large values of $m_0$ and $m_{1/2}$ but which does not always overlap with
the neutralino relic density allowed HB region.  We will leave it to the
reader to assess how much fine-tuning is too much, and also how much
credence one should give to $\Delta_{\rm HS}$ in light of our ignorance
of physics at or around the GUT scale\footnote{Of course, if we take the
mSUGRA model to be the final high scale theory, we would no doubt take
$\Delta_{\rm HS}$ to be our fine-tuning measure, but the judgement to be
made is whether one should treat mSUGRA in this manner.}.

>From a scan over the entire mSUGRA/CMSSM parameter space including LHC
sparticle and Higgs mass constraints, we do find viable regions where EWFT
is at the 1\% level, even for gluino and squark masses well beyond
LHC reach. These regions are characterized by $m_0\sim 8$~TeV and
$A_0\sim \pm 0.6 m_0$. Since these points are spread across a wide range
of $m_{1/2}$ values ranging up to and perhaps beyond 2~TeV, it appears that 
regions of parameter space with EWFT at the 0.5-1\% levels (but with
very large values of $\Delta_{\rm HS}$) will
persist even after the most ambitious LHC SUSY searches are completed.

To conclude, we remind the reader it was the realization that SUSY can
solve the {\em big hierarchy} problem which provided the rationale for
low scale SUSY.  This remains unaltered by LHC and Higgs mass
constraints. The underlying hope was that with sparticles close to the
weak scale, there would be no hierarchy problem. The data seem to
indicate that, at least in the mSUGRA framework, EWFT at the percent
level is mandatory. It is difficult to say whether these considerations
point to the failure of the mSUGRA model, or whether the little
hierarchy is the result of an incomplete understanding of how soft
supersymmetry breaking parameters arise.  While we continue to regard
models with low EWFT as especially interesting, it appears difficult to
unilaterally discard SUSY models that are fine-tuned at a fraction of a
percent or a part per mille, given that these provide the solution of
the much more pressing {\em big hierarchy problem}. 
Our results provide a
quantitative measure for ascertaining whether or not the remaining 
mSUGRA/CMSSM model parameter space is excessively fine-tuned, and so
could provide impetus for considering alternative SUSY models.

%%%%%%%%%%%%%%%%%%%%%%%%%%%%%%%%%%%%%%%%%%%%%%%%
\acknowledgments
%%%%%%%%%%%%%%%%%%%%%%%%%%%%%%%%%%%%%%%%%%%%%%%%
We thank R.~Nevzorov for many discussions about fine-tuning, including
discussions about the potential contributions from first and second
generation of sfermions. We are also grateful to I.~Gogoladze for
raising the issue of the large logs at PHENO 2012, and to C.~Csaki for
clarifying conversations at this meeting.  This work was supported in
part by grants from the U.~S. Department of Energy. 

% ---- Appendix ---- %

% ---- Bibliography ----
%

\end{document}